\documentclass[twocolumn]{svjour3}         %
\smartqed  %
\usepackage{graphicx}
\usepackage{subfigure}
\usepackage{amssymb} %
\usepackage{xspace}
\usepackage{algorithmic}
\usepackage{algorithm}
\usepackage{url}

\newcommand{\hnc}{\emph{Hide\&Crypt}\xspace}
\newcommand{\hns}{\emph{Hide\&Seek}\xspace}

\newcommand{\naive}{\emph{Naive}\xspace}

\newcommand{\Chnc}{\textit{C-Hide\-\&\-Hash}\xspace}
\newcommand{\Chns}{\textit{C-Hide\-\&\-Seek}\xspace}

\newcommand{\chnh}{\textit{C-Hide\-\&\-Hash}\xspace}
\newcommand{\chnc}{\textit{C-Hide\-\&\-Hash}\xspace}
\newcommand{\chns}{\textit{C-Hide\-\&\-Seek}\xspace}

\newcommand{\chncabbr}{\textit{C-H\-\&\-Hash}}
\newcommand{\chnsabbr}{\textit{C-H\-\&\-Seek}}

\newcommand{\maxdist}{\emph{maximum}-\emph{distance}\xspace}
\newcommand{\mindist}{\emph{minimum}-\emph{distance}\xspace}
\newcommand{\mostprox}{\emph{mostly}-\emph{in}-\emph{proximity}\xspace}

\newcommand{\floc}{\texttt{FriendLoc}}
\newcommand{\friendloc}{\texttt{Friend\-Locator}\xspace}

\graphicspath{{fig/}}
\begin{document}

\title{Privacy in geo-social networks: proximity notification with untrusted service providers and curious buddies
}

\author{Sergio Mascetti	\and
        Dario Freni \and 
        Claudio Bettini  \and
		X. Sean Wang \and
        Sushil Jajodia
}

\institute{S. Mascetti, D. Freni, and C. Bettini \at
           DICo, Universit\`a degli Studi di Milano \\
           \email{\{mascetti,freni,bettini\}@dico.unimi.it}     %
           \and
           X.S. Wang \at
           Department of CS, University of Vermont \\
           \email{xywang@cs.uvm.edu}
           \and
           S. Jajodia \at
           CSIS, George Mason University \\
           \email{jajodia@gmu.edu}
}
\date{Received: date / Accepted: date}

\maketitle

\begin{abstract}
  A major feature  of the emerging geo-social
  networks is the ability to notify a user when any of his friends (also called buddies) 
  happens to be geographically in proximity. This \emph{proximity service} is
  usually offered by the network itself or by a third party service
  provider (SP) using location data acquired from the users. This
  paper provides a rigorous theoretical and experimental analysis of  
  the existing solutions for the location privacy problem in proximity services. 
  This is a serious problem for users who do not trust the SP to handle their location data, and
  would only like to release their location information in a generalized form to
  participating buddies.  The paper presents two new protocols providing complete privacy with respect to the SP,
  and controllable privacy with respect to the buddies. The analytical and experimental analysis of the protocols takes into account privacy, service precision, and computation and communication costs, showing the superiority of the new protocols compared to those appeared in the literature to date.
 The proposed protocols have also been tested in a full system implementation of the proximity service. 

\end{abstract}

\section{Introduction}
\label{sec:intro}

A geo-social network is an extension of a social network in which
the geographical positions of participants and of relevant resources
are used to enable new information services.
These networks are mostly motivated by the increased availability of GPS-enabled mobile devices that support both Location-Based Services (LBSs), and easy access to the current social networks.

As in most social networks, each user has a contact list of
\emph{friends}, also called \emph{buddies}. A basic service in
geo-social networks is the {\em proximity service} that alerts the
user when any of her buddies is in the vicinity, possibly enacting
other activities like visualizing the buddy's
position on a map, or
activating a communication session with the buddy.
Such proximity services, often called
\emph{friend finder}, are already available as part of geo-social
networks (e.g., \emph{Brightkite}\footnote{\url{http://brightkite.com}}), as part of a suite of map and navigation
services (e.g., \emph{Google Latitude}\footnote{\url{http://www.google.com/latitude}}), or as an independent service
that can be integrated with social networks (e.g., \emph{Loopt}\footnote{\url{http://www.loopt.com}}).

From a data management point of view, a proximity service involves the
computation of a 
range query over a set of moving
entities issued by a moving user, where the range is a distance
threshold value decided by the user.
All existing services are based on a centralized architecture
in which location updates,
issued from mobile devices,
 are acquired  by the SP, and proximity is computed based
on the acquired locations. 

\subsubsection*{Privacy threats in LBS}
While proximity services are very attractive for many social network
users, the repeated release of information about where the user is at
a given time raises severe privacy concerns.  This is an issue that
has been deeply investigated in the last years for general LBSs, even
if no general consensus has been reached about how
the privacy problem should be defined, measured and, consequently,
alleviated. For this reason we briefly illustrate the general problem
before describing our approach.

The lack of agreement observed in the literature is mainly due to the dual role
that location information plays in LBS privacy.
On one side, location is considered the private data that a user does
not want to disclose, because it may be itself sensitive information,
or because it may lead to disclosure of sensitive information.  For
example, by knowing that a user is in a synagogue during an important
religious ceremony, an adversary may infer, with a certain
probability, the user's religious belief, which may be considered a
privacy violation by this user.
On the other side, location information may act as a
\emph{quasi-identifier}, i.e., when this information is joined with external data
it may compromise the user's anonymity,
and hence allow an adversary to associate the user's identity with the sensitive
information related to the service. For example, suppose a user
subscribes to a location-based dating service using a pseudonym; even
if the locations released to the service are not considered sensitive by her, her
identity can be recovered by first deriving, from her trace of
movements, her home and workplace addresses and then joining these addresses with public data, like a telephone directory. 
In this way, the adversary can deduce the identity of the dating service user, a privacy violation.

Since the specific position of a user at a given time can play either roles illustrated above, two different privacy notions 
have appeared in the LBS literature: a) \emph{location privacy} which
assumes that untrusted parties may know the user's identity but not
the user's location, or at least not the user's precise location, which is considered sensitive and has to be
protected \cite{KidoICPS05,YiuICDE08,MascettiMDM09,JensenPilbaBook09}, and b) \emph{identity privacy} in which the anonymity of
the user must be preserved by avoiding the (precise or imprecise) location information being used as a
quasi-identifier \cite{GruteserMobiSys03,KalnisTKDE07,MascettiJLBS08,GedikIEEETMC08}. Techniques adopted for the second notion (e.g., spatial
cloaking to include at least $k$ users in the released location)
do not necessarily provide privacy guarantees for the first notion, and vice versa. 
In Section \ref{sec:rel} we shortly review the main techniques applicable to proximity services, including approaches trying to address both privacy notions.

\subsubsection*{The location privacy problem in proximity services}

In this paper we consider geo-social networks proximity services in
which a user usually knows the identity of her buddies, or may
easily discover it. 
In this context, identity privacy is not an issue, since the anonymity of the buddies is not assumed. 
For this reason, the problem we address is a \emph{location privacy} preservation, i.e., the first notion accordingly to the above discussion.
We assume that both SP and buddies are considered as \emph{potential adversaries}, that is, 
a) the users do not trust the service provider that will handle their (sensitive) location data, and
b) the users would like to 
control the precision of the location data released
to their buddies.

The above assumption of limited trust is formalized in terms of privacy preferences. Regarding a), we make the strong requirement that SP should not acquire any location information about the users; regarding b), each user can specify the finest precision of the location information that can be disclosed to her buddies, where the precision is in terms of a spatial region of uncertainty containing the current location.
For example, user Alice allows her buddy Bob to be notified when she is in proximity, but she wants a) to hide completely her location to the SP, and b) to ensure that whatever proximity threshold (i.e., the radius of the proximity query) Bob is using, he cannot understand where exactly Alice is located within a region decided by herself (e.g., the whole university campus).

Existing proximity services do not offer any protection regarding
point a) above other than legal privacy policy statements, and they
offer a very limited control regarding point b); for example, some
solutions allow the user to limit the location released to the buddies
to the precision level of city. 
Preliminary studies of this problem have appeared in the academic literature 
\cite{ZhongPET07,MascettiMDM09,SiksnysSSTD09,MascettiSDM09,SiksnysMDM10}, 
and are analysed in detail in Section \ref{sec:rel}, but the solutions provided in these studies have limitations either in terms of safety, system costs, or in terms of flexibility in the specification of user preferences and adversary model.

\subsubsection*{Contribution}
The main contributions of this paper are the following.
\\[2mm]
i) This is the first comprehensive rigorous study of location privacy in proximity services, explicitly taking into account privacy control with respect to buddies.
\\[2mm]
ii)  Two new location privacy preserving protocols are designed, formally analyzed, and empirically
tested, showing their superiority with respect to existing solutions.
\\[2mm]
We formally model the privacy preferences that each user can specify
as well as the properties that a set of messages,
exchanged between a user and the SP,
should have in order to satisfy the privacy preferences.
To the best of our knowledge,
the formal model proposed in this paper
is the first one to consider adversaries having
a-priori probabilistic knowledge of users' location.
For example, it is possible to model the presence of common knowledge, like the fact that 
a user is more likely to be located in the country where she lives
rather than in a foreign one.

The formal model is used to prove that the
protocols proposed in this paper
guarantee location privacy protection. Intuitively, we prove that,
for each possible a-priori probabilistic knowledge of users' location,
each set of messages exchanged between a user and the SP
satisfies the user's privacy preferences.

Our theoretical and experimental study shows that,
in addition to privacy protection,
the proposed protocols offer other two main advantages:
they have sustainable communication and computation costs and
they have low impact on the quality of service.
In order to tackle the problem of system costs, we adopt a centralized architecture.
This solution, 
not only supports
current business models,
but, by reducing the communication and computation
costs on the clients, it is also more appropriate for a proximity service with respect to a decentralized architecture like the ones proposed in previous
works \cite{ZhongPET07,MascettiMDM09}.
For what concerns the quality of service,
the performance of the proposed protocols
can be controlled by user's preferences.
Indeed, differently from previous solutions \cite{ZhongPET07,SiksnysSSTD09},
we allow each user to specify one parameter for the proximity threshold
 and a different parameter for the privacy preference.
In addition to be more flexible than existing approaches,
our techniques provide much higher quality of service.

The two protocols shown in this paper present the same service precision
but differ in the trade-off between privacy guarantees and system costs.
Indeed, the first protocol, called \chns\footnote{\emph{C} stands for \emph{centralized}},
is shown to provide complete protection with respect to the SP,
and to satisfy the privacy requirements of each user with respect to her buddies.
Its efficiency is comparable with the simplistic solution adopted
in current services for proximity computation that provides no privacy protection. 
The second protocol, called \chnc, offers the same guarantees, but provides an even higher level of privacy with respect to the buddies at the cost of higher
communication and computation costs.

The rest of the paper is organized as follows.
In Section~\ref{sec:rel} we discuss related work. In Section~\ref{sec:prob} we describe more formally the problem we are addressing in terms of privacy concerns, privacy requirements, and adversary models. 
In Section~\ref{sec:tec} we illustrate the two proposed protocols,
and in Section~\ref{sec:analysis} we study their formal properties, including the satisfaction of privacy requirements, the computational and communication costs, and the service precision. In Section~\ref{sec:implementation} we describe the system implementation, and in Section~\ref{sec:exres} we report experimental results.
Section~\ref{sec:conc} concludes the paper with a discussion of possible extensions.

\section{Related Work}
\label{sec:rel}
As mentioned in Section~\ref{sec:intro},
there are three main approaches for privacy preservation in LBS
based on whether they deal with a) identity privacy, b) location privacy, 
or c) a combination of these two.
In Section~\ref{sub:relLoc} we first discuss the applicability to our reference scenario of 
the approaches that deal with (b). We discuss the approaches that deal with (c) in Section~\ref{sub:relComb}.
Then, in Section~\ref{sub:relPS}, we focus on the existing contributions that are
specifically related to proximity services.
An extensive survey of privacy preserving techniques in LBS can be found in \cite{PilbaBook}.

For what concerns approaches that deal with (a), i.e., identity privacy only,
the main technique is based on the application of the $k$-anonymity principle \cite{SamaratiTKDE01} to LBS, and was proposed by Gruteser et al. \cite{GruteserMobiSys03}.
$k$-anonymity is achieved by ensuring that the user's location sent to the SP as part of the request is a region sufficiently large to include, in addition to the issuer, other $k-1$ potential issuers.
Several variants of this idea and different algorithms for spatial cloaking have been proposed. For example, Gedik et at.
\cite{GedikIEEETMC08} illustrate a cloaking algorithm allowing each user to choose a personalized value of $k$,
while others propose algorithms proved to be safe also in case
the adversary knows the defense function \cite{KalnisTKDE07,MascettiJLBS08}.
As mentioned in Section~\ref{sec:intro}, 
these techniques are not applicable in the reference scenario considered in this paper, since we consider services in which users may not be anonymous, independently from location information.
In Section~\ref{sec:conc} we briefly discuss
how the techniques we propose in this paper
can be extended to provide identity privacy for those proximity services that require anonymity.

\subsection{Location privacy protection}
\label{sub:relLoc}
The intuition behind location privacy (i.e., the first privacy notion given in Section~\ref{sec:intro})
is that users perceive their location
as private information.
However, they may tolerate that some location information
is disclosed if it is sufficiently unlikely
that the adversary discovers their \emph{precise} location.
To achieve this result, techniques based on different ideas
have been proposed.

One idea is to send requests from fake locations together with
the request from the real user's location (e.g., \cite{KidoICPS05}).
The main problem with the techniques implementing this idea 
is that a large number of fake request is necessary in order to
guarantee privacy protection, while the system costs grow linearly in the number of fake requests.

Another solution consists in sending a request
(e.g., a $K$-NN query) from a fake location
and incrementally retrieve results (e.g., NN resources) from the SP
until the client can reconstruct the result to the query
centered in the real user's location \cite{YiuICDE08}.
Privacy is guaranteed because
the SP can only discover that the user is located
within a region
without learning the exact location.
The distance between the real user's location
and the fake location used in the request
determines
a trade-off between
privacy and performance.
Indeed, if the distance is large,
the size of region discovered by the SP is also large,
but this results in high system costs.
These techniques have been applied mostly for LBS performing
$k$-NN spatial queries, and do not apply to proximity detection.

A third family of techniques to enforce location privacy is based on the idea
of enlarging the user's precise location before it is sent to the SP to a
\emph{generalized region} in order to decrease its sensitivity  (among
others, 
\cite{MascettiMDM09,SiksnysMDM10,GhinitaGIS09}).
Some of these techniques are specifically designed for proximity
services and we discuss them in details in Section~\ref{sub:relPS}.
The main technical problem 
is how to process spatial queries
in which the parameters are generalized regions
instead of exact locations.
On the other hand, the advantage is that
the generalized region can be specified
as a user preference
before any information is sent by the client.
Indeed, this is the solution we adopt in this paper to protect
a user's privacy with respect to her buddies.
We actually prove that when a user specifies a generalized region,
her buddies do not acquire any location information
about that user, except the fact that she
is inside the generalized region.

\subsection{Identity and location privacy protection combined}
\label{sub:relComb}
Some of the solutions proposed in the literature
aim at providing both location privacy and identity privacy.
For example, in Casper \cite{MokbelVLDB06},
users' locations are generalized
to a region that contains at least $k$ users and
that has an area
not smaller than a user-specified threshold.
The problem with this solution is that it is insecure
in case the adversary knows the generalization technique.
Another similar technique based on that of  \cite{YiuICDE08} is reported in 
\cite{WangPALMS09} to tackle both privacy notions with one algorithm.

Other solutions providing both
location and identity privacy
are inspired
to private information retrieval (PIR) methods.
The idea is to encrypt
the information exchanged with the SP,
and to process the corresponding query in an encrypted form,
so that no location information is revealed to the SP.
The techniques proposed in \cite{GhinitaSIGMOD08,PapadopoulosVLDB10}
are specifically designed for NN queries,
while \cite{KhoshgozaranKAIS10} considers range queries over static resources, which is still not the proper setting for proximity detection.
Khoshgozaran et al.~\cite{KhoshgozaranSIN09} propose a
system to maintain an encrypted index on the server side and
efficiently update it, which makes it suitable for maintaining a
database of moving buddies. The system supports encrypted range and
$k$-NN spatial queries, hence it could be used to offer proximity
based services. However, the system requires users to be organized in
groups, with each group sharing a symmetric secret key, and all the
users in a group must trust each other.
Furthermore, the proposed
techniques for retrieving the query results seem to be vulnerable to
cardinality attacks ~\cite{PapadopoulosVLDB10}, if the SP has
a-priori knowledge about the distribution of the users.

These \emph{encryption-based} techniques guarantee location and identity privacy
because no location information is disclosed.
Consequently, if it is assumed that users are anonymous,
then identity privacy is also guaranteed, since no location information
can be used to re-identify the user.
Considering location privacy,
these techniques provide the same protection
as a solution based on location-enlargement 
in which the user's location is generalized to the entire world,
i.e., they provide the maximum privacy protection.
For this reason, in this paper we adopt this solution
to guarantee privacy with respect to the SP.

\subsection{Location privacy protection in proximity services}
\label{sub:relPS}
Computing proximity involves the continuous evaluation
of spatial range queries over a set of moving entities,
with a dynamic radius range \cite{ChonAA03,Saltenis-SIGMOD00}.
The literature on this problem is both from the database,
and the mobile computing community;
recent contributions are briefly surveyed in \cite{MyllymakiPMC07},
where an efficient algorithm for proximity detection
named \emph{Strips} is presented.
The goal of this and similar approaches (e.g., \cite{YiuVLDB10})
is the efficiency
in terms of computation and communication complexity,
while privacy issues are mostly ignored.

Ruppel et al. \cite{RuppelTKL06} propose a technique for privacy
preserving proximity computation based on the application of a
distance preserving transformation on the location of the
users. The problem with this solution is that the SP is able to obtain the exact distances between
users, and this can lead to a privacy violation. For example, by using
this knowledge, it is possible to construct a weighted graph of all
the users, assigning to each edge connecting two users their exact
distance. It is easily seen that a ``relative''
distribution of the user locations can be extracted from this graph. If the SP
has a-priori knowledge about the distribution of the users (as considered in our paper), it
is possible to combine the distribution resulting from the graph
with the a-priori one, thus revealing some location information about
the individuals.  In addition, there is no privacy guarantee with
respect to the other users participating in the service.  The
solutions we propose in this paper do not reveal to the SP any
information about the distance between users, and let users define the
privacy requirement about the location information that
buddies can acquire.

Zhong et al. propose three different techniques
for privacy preservation in proximity-based services
called \emph{Louis}, \emph{Lester} and \emph{Pierre} \cite{ZhongPET07}.
These techniques are decentralized secure computation protocols
based on public-key cryptography. 
\emph{Louis} is a three-parties secure computation protocol.
By running this protocol, a user $A$ gets to know whether
another user $B$ is in proximity 
without disclosing any other location information to $B$ or to 
the third party $T$ involved in the protocol. $T$ only helps 
$A$ and $B$ compute their proximity, and it is assumed to 
follow the protocol and not to collude with $A$ or $B$. 
However, $T$ learns whether $A$ and $B$ are in proximity.
Considering our adversary model, which will be
explained in detail in Section 3.3, this third party 
cannot be the SP that may use proximity information 
to violate location privacy, and it is unlikely 
to be played by a third buddy since it would 
involve significant resources.
The \emph{Lester} protocol allows a user $A$
to compute the exact distance from a user $B$
only if the distance between the two users
is under a certain threshold chosen by $B$.
The main advantage of these two techniques is that
they protect a user's privacy without introducing
any approximation in the computation of the proximity.
However, \emph{Louis} incurs in significant
communication overheads, and \emph{Lester}
in high computational costs.
In addition, the only form of supported privacy protection
with respect to the buddies
is the possibility for a user to refuse to
participate in the protocol initiated
by a buddy if she considers the requested proximity
threshold too small.
The \emph{Pierre} protocol partitions
the plane where the service is provided
into a grid, with each cell having edge
equal to the requested distance threshold. 
The locations of
the users are then generalized to the
corresponding cell, and two users are considered
in proximity if they are located in
the same cell or in two adjacent cells. The
achieved quality of service decreases as the requested proximity threshold grows.
We will explain in more
detail the actual impact on service precision in Section~\ref{sec:exres}. 
Finally, it should be 
observed that Lester and Pierre protocols
are based on a buddy-to-buddy communication, 
and although this can guarantee total
privacy with respect to the SP (as no SP is
involved in the computation), scalability issues
may arise 
since each time a user moves
she needs to communicate her new position
to each of her buddies.

Another solution for privacy preserving computation
of proximity, called \friendloc, has been proposed by 
\v{S}ik\v{s}nys et al. \cite{SiksnysSSTD09}.
Similarly to Pierre, two users are considered in proximity
when they are located in the same cell or two adjacent 
cells of the grid constructed considering the
proximity threshold shared by the users. An interesting
aspect of the proposed solution is the location update 
strategy, which is designed to reduce the total number
of location updates to be sent by the users, hence reducing
communication costs. Two users share a hierarchy of
grids, where a \emph{level} identifies each grid.
The larger the value of the level is, the finer the grid.
The highest level grid is the one in which the edge
of a cell is equal to the proximity threshold.
The detection of proximity is then incremental, i.e.
if two users are in adjacent cells at the level $n$ grid, 
then their respective cells in the grid of level $n+1$ are checked,
until they are detected either not to be in proximity, or to be
in proximity considering the highest level grid. With this
solution, when two users are detected not to be in proximity
at a certain level $l$, there is no need for them to check
again the proximity until one of them moves to a different cell of the level $l$ grid. 
As a consequence, less location updates are needed, and this is experimentally
shown to significantly reduce the total number of messages exchanged.
However, the \friendloc
protocol reveals some approximate information about the 
distance of users to the SP (e.g. the level in which the incremental 
proximity detection protocol terminates and whether the
buddies are in proximity at that level). 
As already observed for the Louis protocol, in our adversary model this
information can lead to a privacy violation.
Furthermore,
the impact on the quality of service of using a large proximity threshold 
is identical to the Pierre protocol discussed above.

A more recent solution by the same authors \cite{SiksnysMDM10}, called
\texttt{VicinityLocator}, solves this problem by letting users specify their privacy
preferences as spatial granularities (see Section~\ref{sub:prefs}),
independently from the requested proximity threshold. A similar
location update strategy is employed to minimize the communication
costs. However, similarly to \friendloc, the SP learns some
information about the distance among the users, and this could lead to a
privacy violation in our adversary model.

In previous work \cite{MascettiSDM09,MascettiMDM09}, we proposed
different protocols for preserving privacy in proximity services. The
\emph{Longitude} solution \cite{MascettiSDM09} translates the
considered space to a toroid, and a distance preserving transformation
is applied to the locations of users. The SP participates in a form of
three party secure computation of proximity, leading to an approximate
but quite accurate service precision, guaranteeing 
privacy
requirements with respect to buddies similar to the ones presented in
this paper.  Longitude also guarantees complete privacy with respect
to the SP under the assumption that he has no a-priori knowledge on
the distribution of users, i.e., when a uniform distribution is
assumed.  In this paper we defend also against SP having arbitrary
a-priori distributions, showing that by running our protocols they
don't acquire any additional location information.
The \hns and \hnc protocols \cite{MascettiMDM09} are hybrid techniques
in which the SP performs an initial computation of the proximity. In some cases, the SP is not able to decide whether two users are in proximity, and a buddy-to-buddy protocol is triggered. An important
difference with respect to the protocols we are presenting here is that the
SP is not totally untrusted: users can specify a level of location
precision to be released to the SP and (a different one) for
buddies. This hybrid approach significantly reduces communication
costs with respect to decentralized solutions when privacy
requirements with respect to the SP are not too strict.

\section{Problem formalization}
\label{sec:prob}
In this section we formally define the service we are considering, the
users' privacy concerns and requirements, the adversary model, and
the occurrence of a privacy violation.

\subsection{The proximity service}
\label{sub:service}
By issuing a proximity request, user $A$ is interested to know, for
each of her buddies $B$, if the  following condition is satisfied:
\begin{equation}
\label{eq:distance-threshold}
 d(loc_{A}, loc_{B})\le \delta_A
 \end{equation}
where $d(loc_{A}, loc_{B})$ denotes the Euclidean 
distance
between the reported locations of $A$ and $B$ and 
$\delta_{A}$ is a threshold value given by $A$. 
When (\ref{eq:distance-threshold}) is true, we say
that \emph{$B$ is in the proximity of $A$}.  
The proximity relation is not symmetric, since $\delta_B$ may be different from  $\delta_A$,

In this paper we consider services in which the buddies of a
user are pre-determined.
We call these services ``contact-list-based'',
since buddies are explicitly added as ``friends'', like in most social
networks and instant messaging applications.
This is in contrast to ``query-driven'' proximity services, in which
buddies can be retrieved through a query based,
for example, on the interests of the buddies.
Technically, the main difference is that in the ``contact-list-based''
service it is reasonable to assume that each user can share a secret
with each of her buddies, as we do in our proposed techniques.
On the contrary, in the case of ``query-driven''
services, the set of buddies may change dynamically, and
the number of buddies can be potentially very large. 
In this situation,
it may not be practical to share a secret with each buddy.

With the presence of a service provider (SP),
and in absence of privacy concerns,
a simple protocol can be devised to implement the proximity service:
The SP receives location updates from each user 
and stores their last known positions,
as well as the distance threshold $\delta_A$ for each user $A$.
While in theory each user can define different
threshold values
for different buddies, in this paper, for simplicity,
we consider the case in which each user $A$ defines a single value $\delta_A$
for detecting the proximity of all of her buddies.
When the SP receives a location update,
it can recompute the distance between $A$ and each
buddy (possibly with some filtering/indexing strategy for efficiency)
and communicate the result to $A$.
In a typical scenario, if $B$ is in proximity,
$A$ may contact him directly or through the SP;
however, for the purpose of this paper,
we do not concern ourselves as what $A$ will do once notified.
In the following of this paper we refer to the above protocol
as the \naive protocol.

\subsection{Privacy concerns and  privacy requirements}
\label{sub:prefs}

The privacy we are considering in this paper
is \emph{location privacy}: we assume that
a user is concerned about the uncontrolled disclosure
of her location information at specific times.

Considering the \naive protocol,
it is easily seen that the SP obtains 
the exact location of a user each time she issues a location update. 
Furthermore, a user's location information is also disclosed to her buddies. 
If Alice is in the proximity of Bob (one of her buddies),
then Bob discovers that Alice is located in the
circle centered in his location with radius $\delta_{Bob}$.
Since $\delta_{Bob}$ is chosen by Bob and can be set arbitrarily
without consent from Alice,
Alice has no control on the location information disclosed to Bob.

Our definition of location privacy is based on the idea that
the users should be able to control the location information to be disclosed.
In the considered services, a user may prefer the service
provider to have as little information about her location as possible,
and the buddies not to know her exact position,
even when the proximity is known to them. 
Moreover, the exchanged information should be protected from any eavesdropper.

In general, the level of location privacy
can be represented by the uncertainty
that an external entity has about the position of the user.
This uncertainty is a geographic region,
called \emph{minimal uncertainty region} (MUR),
and its intuitive semantics is the following:
the user accepts that the adversary knows
she is located in a MUR $R$, but no information
should be disclosed about her position
within $R$.

In the solution proposed in this paper,
each user can express her privacy preferences
by specifying
a partition of the geographical space defining
the MURs that she
wants guaranteed.
For example, Alice specifies that
her buddies
should never
be able to find out the specific campus building 
where Alice currently is;
in this case, the entire campus area is the minimal uncertainty region. 
The totality of these uncertainty regions for a user
can be formally captured with the notion of \emph{spatial granularity}. 

While there does not exist a formal definition of spatial granularity
that is widely accepted by the research community,
the idea behind this concept is simple.
Similar to a temporal granularity \cite{BJW:book},
a spatial granularity can be considered a subdivision of the spatial domain
into a discrete number of non-overlapping regions, called \emph{granules}. %
In this paper, for simplicity, we consider only
granularities\footnote{Here and in the following,
when no confusion arises, we use the term ``granularity''
to mean ``spatial granularity''.}
that partition the spatial domain, i.e.,
the granules of a granularity do not intersect and the union of all the granules in a granularity
yields exactly the whole spatial domain. 
Each granule of a granularity $G$ is identified by an \emph{index} (or a \emph{label}).
We denote with $G(i)$ the granule of the granularity $G$
with index $i$.

Users specify their {\em privacy requirements}
via spatial granularities,
with each granule being a MUR. 
The two extreme cases in which a user requires no privacy protection and
maximum privacy protection, respectively, can be naturally modeled.
In one extreme case, if a user $A$ does not want her privacy
to be protected
then $A$ sets her privacy preference to the \emph{bottom granularity} $\bot$
(a granularity that contains a granule for each basic element,
or pixel, of the spatial domain).
In the other extreme, if user $A$
wants {\em complete} location privacy
then she sets her privacy preference to the \emph{top granularity} $\top$, i.e.,
the granularity that has a single granule covering the entire spatial domain. In this case, $A$
wants the entire spatial domain as MUR.

In the following of this paper, we assume that each user $A$
specifies a granularity $G_A$
defining her location privacy requirements with respect to all buddies. 
Our approach can be easily extended to model the case in which
a user specifies a different granularity for
a different buddy or for a different group of buddies,
as discussed in Section~\ref{sec:conc}.
We also assume that each user's privacy requirement 
with respect to the SP is the entire spatial domain,
i.e., the user does not want to disclose any location information
to the SP.

\subsection{Adversary model and privacy preservation}
\label{sub:advmodel}

We consider two adversary models, for the SP and the buddies, respectively.
Assuming the SP and the buddies as potential adversaries,
also models other types of adversaries.
Firstly, it models the case of an external entity taking control
of the SP system or of a buddy's system.
Secondly, it models the case of an external entity
eavesdropping one or more communication channels between users and the SP.
Note that, in the worst case, the eavesdropper can observe all the messages
that are exchanged in the protocol.
Since the same holds for the SP,
the eavesdropper can learn at most what the SP learns.
Since in this paper we prove  that
the SP does not acquire any location information, then
 the same holds for an eavesdropping adversary.

The techniques we present in this paper not only guarantee
each user's privacy requirement against these two adversary models, but
also in the case of a set of colluding buddies.
In Section~\ref{sub:collusion} we also discuss which privacy guarantees
are provided by our techniques in case one or more buddies collude with the SP.

In both adversary models
we assume that the adversary knows:
\begin{itemize}
\item the protocol,
\item the spatial granularities adopted by each user, and
\item an a-priori probabilistic distribution of the locations of the users.
\end{itemize}
The two models differ in
the sets of messages received during a protocol
run, and in their ability (defined by the protocol in terms of
availability of cryptographic keys) to decrypt the content of the
messages.
 
The a-priori knowledge of the location of a user $A$ is given by a location random 
variable 
$pri_{A}$ with the probability mass distribution denoted $P(pri_{A})$.
In other words, as prior knowledge we assume
that the location of a user $A$ follows a known distribution given by the
distribution of the random variable $pri_{A}$. 
Note that in this paper
we assume the spatial domain
is discrete, i.e., a countable set of ``pixels''.

Let $M$ be the set of messages
exchanged between the entities involved in the service.
The adversary can compute the \emph{a-posteriori} probability
distribution of the location random variable $post_{A}$ as the distribution of the location of $A$ under the given messages $M$ and
the prior knowledge $pri_{A}$:
\[  P(post_{A})=P( loc_{A} | M, pri_{A}) \]
Technically, we may view $loc_{A}$ as 
a uniform random variable over the spatial domain, i.e., the
possible location of $A$ when no knowledge is available.

The condition for privacy preservation is
formally captured by Definition~\ref{def:MUR}.
\begin{definition}
\label{def:MUR}
Given a user $A$ with privacy requirement $G_A$, and
$M$ the set of messages exchanged by the proximity service
protocol in which $A$ is participating, $A$'s privacy requirement is said to
be {\em satisfied} if
\[
P(loc_{A} | M, pri_{A}, loc_{A}\in g_A) = P(loc_{A} | pri_{A}, loc_{A}\in g_A) 
\]
 for all a-priori knowledge $pri_{A}$ and all granule $g_{A}$ of $G_{A}$.
\end{definition}

The above definition requires that the location distribution of user $A$ does
not change due to the messages $M$, given the a-priori knowledge and the
fact that $A$ is located in $g_{A}$.
Hence, a privacy violation occurs when the adversary acquires,
through the analysis of the protocol messages,
\emph{more} information about the location of $A$ than allowed by her privacy requirements, i.e., when the probability distribution
of the position of $A$ within the region defined by granule $g_A$ changes
with respect to $pri_{A}$.

Since we aim at complete location privacy with respect to
the SP, we use $g_A$ to be the entire spatial domain in the above definition when the SP is concerned. In this case, the definition requires $P(loc_{A} | M, pri_{A}) = P(loc_{A} | pri_{A})$, i.e., $P(post_{A}) = P(pri_{A})$ or {\em no new location information} for each user $A$. In this case, we also say that $A$'s privacy requirement is {\em satisfied with respect to the SP}. For the buddies, user $A$ uses a granularity $G_A$, which may not be $\top$. In this case, the definition requires that with the additional knowledge of $A$ being in a granule, the buddies cannot derive anything more (e.g., where within the granule) from the messages exchanged. In this case, we also say that $A$'s privacy requirement is {\em satisfied with respect to the buddies}.

\section{Defense techniques}
\label{sec:tec}
In this section we present two protocols
to preserve location privacy in proximity-based services.
The protocols are called \chns and \chnc
and they guarantee
privacy protection of a user $A$ with respect to both the SP
and the buddies of $A$.

In order to ensure user's privacy,
the two protocols adopt symmetric encryption techniques.
In the following, we assume that
each user $A$ has a key $K_A$ that is shared with
all of her buddies and is kept secret to everybody else.
Hence, each user $A$ knows her own key $K_A$
and one key $K_B$ for each buddy $B$.
Since we are considering a contact-list-based service, this key exchange is assumed to be performed with any secure method before running our protocols.

For the sake of presentation, we decompose each protocol into
two parts: the \emph{location update} sub-protocol is used by a user to provide her location information,
while the \emph{proximity request} sub-protocol is used by a user to compute the proximity of her buddies.
The location update sub-protocol is almost the same in both of our proposed solutions, and it is presented in Section~\ref{sub:locupd}.
What really distinguishes \chns and \chnc is the proximity request sub-protocol, and this is described in Sections~\ref{sec:hns} and \ref{sub:hnc}, respectively.
We conclude this section with a discussion
about possible technical extensions.

\subsection{The location update sub-protocol}
\label{sub:locupd}
The location update sub-protocol is run by a user
to provide location information to the SP.
In particular, it defines how a user $A$ provides to the SP
the encrypted index of the granule of $G_A$ where she is located.

Before describing the sub-protocol,
we first discuss when it should be run.
Consider the following naive policy:
a user $A$ updates her location only when she crosses
the boundary between two granules of $G_A$, reporting the index of the new granule.
It is easily seen that, independently from
how the location update is performed,
each time this message is received, the
adversary learns that $A$ is
very close to the border between two granules,
excluding many other locations, and hence violating the 
privacy requirements.
Intuitively, the problem of the above policy is that the
probability that a location update is performed at a given time
depends on the location from where the message is sent.

\begin{figure}[htb]%
\centering
\includegraphics[scale=0.4]{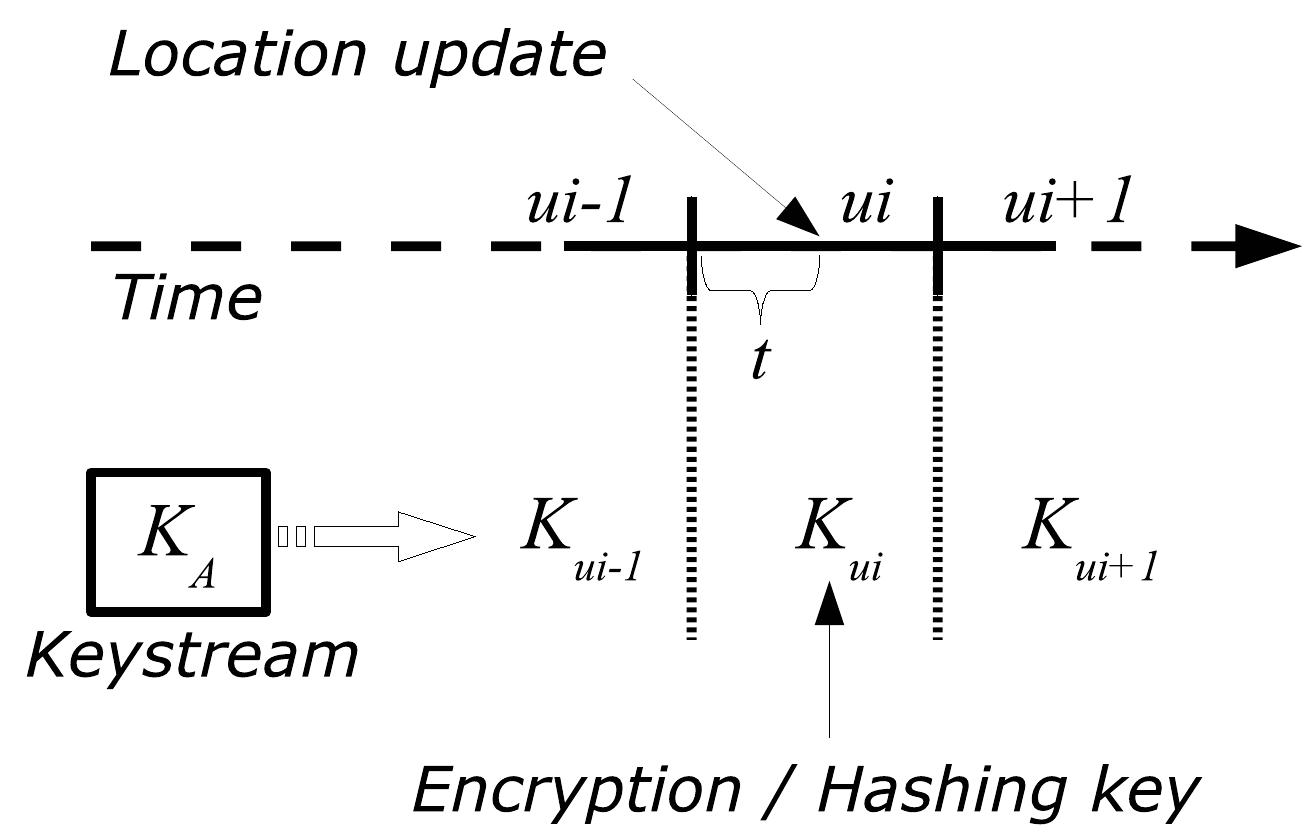}
\caption{Location update policy and generation of single-use keys.}
\label{fig:TimeAndKey}
\end{figure}

The solution we propose is the following:
time is partitioned into \emph{update intervals} and an approximate
synchronization on these intervals among the participating nodes is
assumed.\footnote{In our current implementation, all the messages sent from the SP to
the users contain the timestamp of the SP, allowing clients to synchronize their clocks using a Lamport-style algorithm. The overhead due to this solution is negligible. Other forms of global clock synchronization could also be used as, e.g., using GPS devices.}  
Each update interval has
the same duration $T$ and is identified by an index. Each user has a
value $t$ in $[0, T)$ and sends exactly one location update during
each update interval after that time $t$ elapses from the beginning of
the interval
(see Figure~\ref{fig:TimeAndKey}).
It is easily seen that, by using this update policy, the
location updates are issued independently from the location of the
users.

We now describe how the location update sub-protocol works.
User $A$ first computes the index $i$ of the granule of $G_A$ where she is located.
Then, $A$ encrypts $i$ using a slightly different technique in the two proposed solutions.
In the \chns protocol a symmetric encryption function $E$ is applied, 
while in the \chnc protocol a hashing function $H$ is used.
When applying the hashing function $H$, in order to prevent brute-force attacks,
a secret key is used as a ``salt'', i.e., a secret key is concatenated to $i$, and
the resulting value is given as input to $H$.
In the following, we refer to this salt as the ``key'' used to hash $i$,
and we denote with $H_K(i)$ the hashing of the value $i$ with key $K$.

\begin{figure}[htb]%
\centering
\includegraphics[scale=0.45]{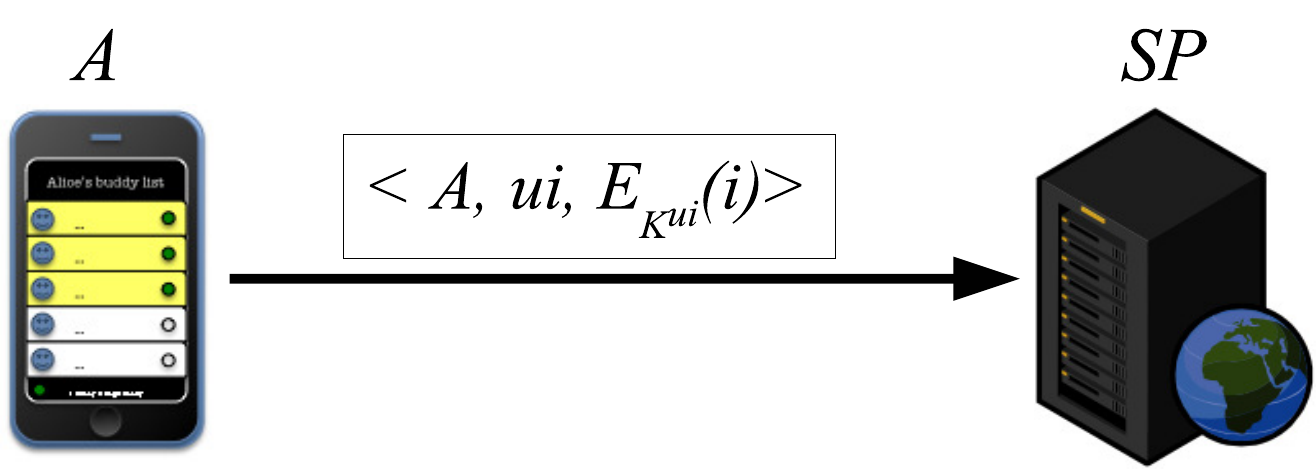}
\caption{Location update sub-protocol in \chns.}
\label{fig:LocationUpdate}
\end{figure}

The safety of the protocols depends on the fact that
the key used to encrypt or hash $i$ is changed at every use.
At the same time, we need the key to be shared
by a user with all of her buddies.
While other techniques can be adopted to achieve this result,
our solution is the following:
the key $K_A$ that $A$ shares with all of her buddies
is used to initialize a keystream.
When user $A$ issues a location update,
she computes the key $K^{ui}$
as the $ui$-th value of this keystream,
where $ui$ is the index of the current update interval
(see Figure~\ref{fig:TimeAndKey}).
Since each user issues a single location update
during each time interval, this solution ensures
that every message is encrypted or hashed with a different key.
Finally, $A$ sends to the SP the message $\langle A, ui, E_{K^{ui}}(i) \rangle$ if running \chns,
and $\langle A, ui, H_{K^{ui}}(i) \rangle$ if running \chnc.
The SP stores this information as the last known encrypted location for $A$.
Figure~\ref{fig:LocationUpdate} shows the message sent from $A$ to the SP
by the \chns protocol.

\subsection{Proximity request with \chns}
\label{sec:hns}

\begin{figure}[htb]%
\centering
\includegraphics[scale=0.45]{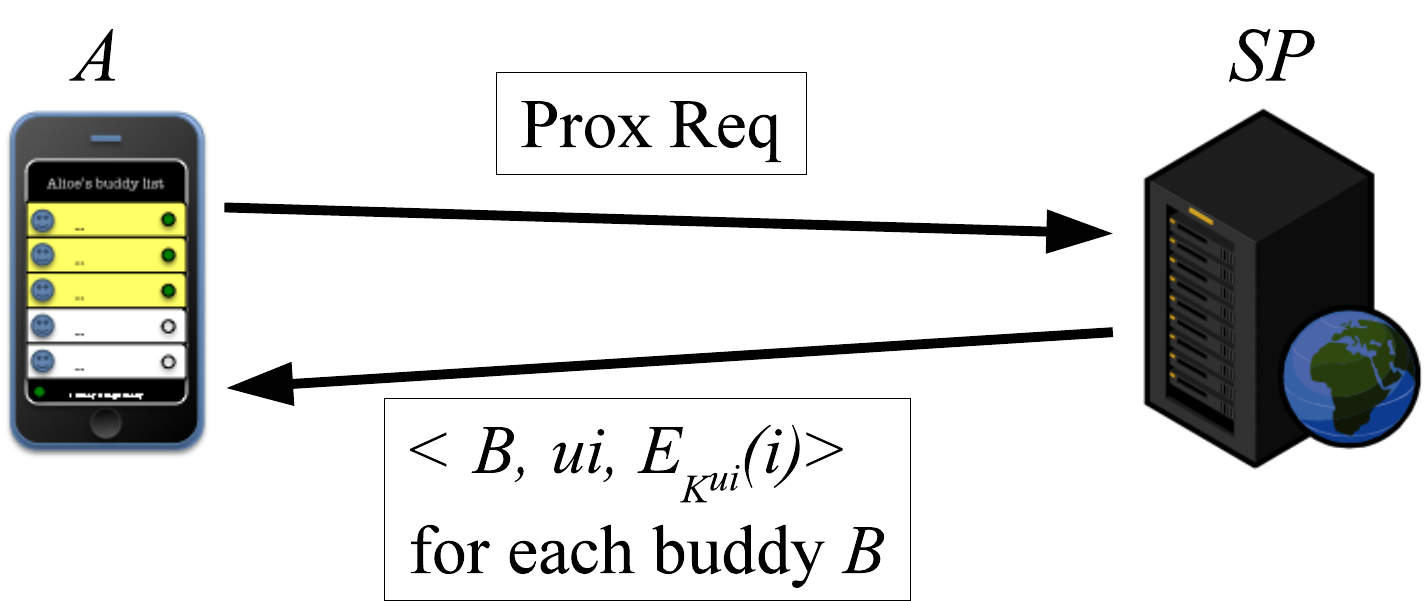}
\caption{Proximity request sub-protocol in \chns.}
\label{fig:CHideAndSeek}
\end{figure}

The proximity request sub-protocol is run by a user
that wants to discover which of her buddies are in proximity.
In the \chns protocol,
this sub-protocol works as follows:
When $A$ wants to discover which buddies are in proximity,
she sends a request to the SP.
The SP replies with a message containing
the last known encrypted location of each buddy of $A$.
That is, for each buddy $B$, $A$ receives a tuple $\langle B, ui, E_{K^{ui}}(i) \rangle$.
Since $A$ knows $K_B$ and the index $ui$ is in the message,
she can compute the value $K^{ui}$
used by $B$ to encrypt his location, and hence she can decrypt $E_{K^{ui}}(i)$.
Finally, since $A$ also knows $G_B$, by using $i$,
she obtains the granule $g_B=G_B(i)$ where $B$ is located.
$A$ can then compute the distance between her exact location
and $g_B$, and compare it with $\delta_A$, finally determining the proximity.
Figure~\ref{fig:CHideAndSeek} shows a graphical representation of the sub-protocol.

Note that we are now considering the proximity between a point and a
region. In this section, we consider that a point and a region are in
proximity, with respect to a distance threshold, if the \emph{minimum}
distance between the two objects is less than the threshold.  Since,
in our protocol, the region represents the area where a user $B$ is
possibly located, this interpretation of proximity means that there is
a possibility for users $A$ and $B$ to actually be in proximity.  The
same \emph{minimum} distance interpretation has been used in related
work on privacy-aware proximity computation. Alternative
interpretations and their effects are discussed in
Section~\ref{sub:qos}.

The \chns protocol provides a simple and efficient solution that, as will be shown in Section \ref{sec:analysis}, completely hides the location of the users to the SP,
and that also guarantees the 
privacy requirements
with respect to the buddies.
However, it reveals exactly the maximum tolerable amount
of location information ($g_B$ for user $B$) to any buddy issuing a proximity request.
Even if their 
privacy requirements 
are guaranteed, users would probably prefer to disclose
as little information as possible about their location when not strictly needed.
For example, is there an alternative solution that does not reveal to a user $A$ the granule information of a buddy $B$ if he is not in proximity?

In the next section we present the \chnc protocol that
provide such a solution and, in general, ensures a higher level of
privacy.  This is achieved at the cost of higher computation and
communication costs, as explained in Section~\ref{sub:costs}.

\subsection{Proximity request in \chnc}
\label{sub:hnc}

The \chnc protocol has two main differences with respect to
\chns.  
The first difference is that a hash function $H$ is
used during the location update, instead of the encryption function.
This is due to the requirement in this protocol to avoid revealing the
relationship between two plaintext values (the granule indexes) by
observing the relationship among the corresponding encrypted values
(see Section~\ref{sec:analysis} for a more detailed
explanation). Since in this protocol we do not need to decrypt the
result of the function, but we only need to check for equality of encrypted
values, hashing can be used.
As specified in Section \ref{sub:locupd}, each location update in \chnc
from user $A$ to the SP is a message containing the tuple $\langle A, ui, H_{K^{ui}}(i) \rangle$.

\begin{figure}[htb]%
\centering
\includegraphics[scale=0.55]{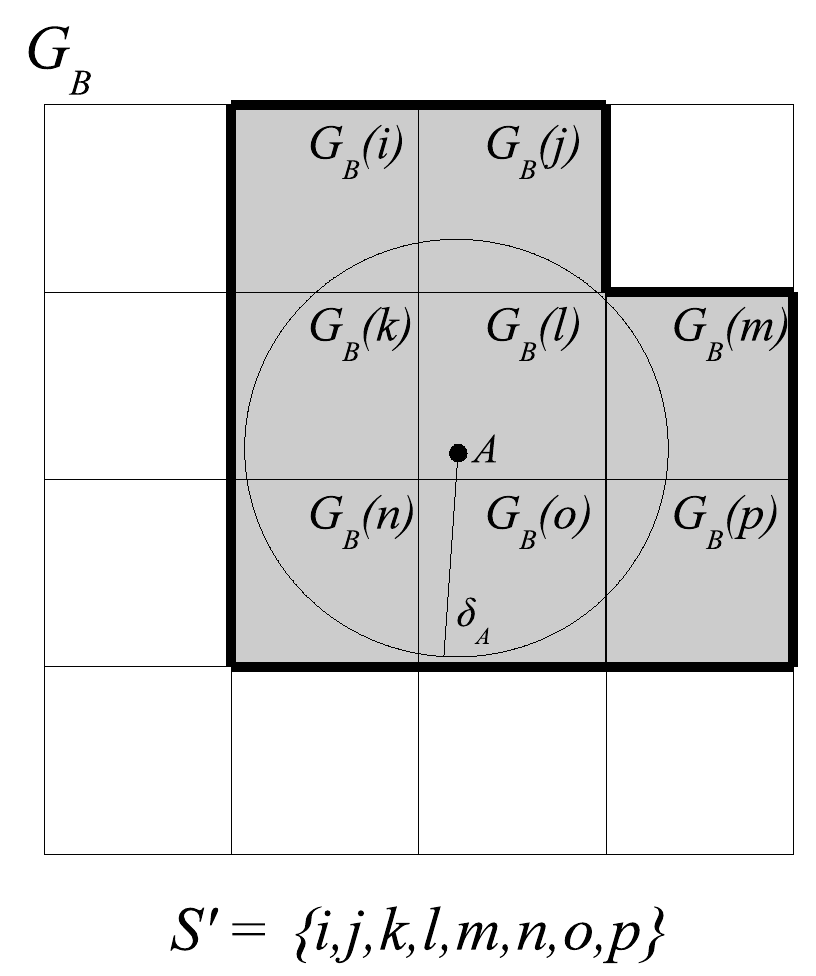}
\caption{Computation of granules of $G_B$ considered in proximity by $A$}
\label{fig:hncintersgran}
\end{figure}

The second and main difference with respect to \chns is the
computation of the proximity request sub-protocol.
The intuition is that when $A$ issues
a proximity request, 
she computes, for each of her buddies $B$, the set of 
indexes of 
granules of $G_B$
such that, if $B$ is located in any granule of the set,
then $B$ is in proximity
(see Figure~\ref{fig:hncintersgran}).
Then, if $B$ provides the granule in which he is located, it is possible to reduce the proximity
problem to the set-inclusion problem,
by checking if that granule is included in the set computed by $A$.
We want to do this set inclusion without revealing to $A$ which of the candidate granules actually matched the granule of $B$.  

More precisely, the computation of a proximity request in the
\chnc protocol works as follows.
When a user $A$ issues a proximity request,
she starts a two-party set inclusion protocol
with the SP.
The protocol is a secure computation,
and consequently 
the SP does not learn whether $A$ is in proximity
with her buddies, and 
$A$ only learns, for each of her buddies $B$,
whether $B$ is in proximity or not,
without learning in which granule $B$ is located.
The secure computation exploits
a commutative encryption function $C$.
In addition to the keys used in the \chns protocol,
at each proximity request, the requesting user and the SP each generates a random key that is not shared with anyone else.
We denote these keys $K_1$ for user $A$ and $K_2$ for the SP.

\floatname{algorithm}{Protocol}
\begin{algorithm}[ht]
\caption{\Chnc : proximity request}
\label{prot:hnc}
\textbf{Input}:
User $A$ knows, the last completed update interval, and the proximity threshold $\delta_A$. Also, for each of her buddy $B$, $A$ knows the granularity $G_B$, the key $K_{B}$ and the value of $sMax(G_{B}, \delta_{A})$. 

\textbf{Protocol}:

(i) Client request from $A$

\begin{algorithmic}[1]
\STATE $proxReq = \emptyset$
\STATE generate a random key $K_1$
\FOR{each buddy $B$ of $A$}
\STATE $S' = \{j \in \mathbb{N} \mbox{ s.t. } minDist(loc_{A}, G_B(j)) \leq \delta_{A} \}$
\STATE $S'' =$ a set of $sMax({G_B}, \delta_A ) - |S'|$ non-valid random indexes.
\STATE $S = S' \cup S''$
\STATE $K^{ui}$ is the $ui$-th value of the keystream initialized with $K_{B}$
\STATE $ES=\bigcup_{i \in S} C_{K_1}(H_{K^{ui}}(i))$
\STATE insert $\langle B, ui, ES \rangle$ in $proxReq$
\ENDFOR
\STATE $A$ sends $proxReq$ to the SP
\end{algorithmic}

(ii) SP response

\begin{algorithmic}[1]
\STATE $proxResp = \emptyset$
\STATE generate a random key $K_2$
\FOR{each $\langle B, ui, ES \rangle$ in $proxReq$}
\STATE $ES' = \bigcup_{e \in ES}C_{K_2}(e)$
\STATE retrieve $\langle B, ui, h_B \rangle$ updated by $B$ at update interval $ui$
\STATE $h' = C_{K_2}(h_B)$
\STATE insert $\langle B, ES', h' \rangle$ in $proxResp$
\ENDFOR
\STATE SP sends $proxResp$ to $A$
\end{algorithmic}

(iii) Client result computation

\begin{algorithmic}[1]

\FOR{each $\langle B, ES', h' \rangle$ in $proxResp$}
\STATE $h'' = C_{K_1}(h')$
\IF{$h'' \in ES'$}
\STATE $A$ returns ``$B$ is in proximity''
\ELSE
\STATE $A$ returns ``$B$ is not in proximity''
\ENDIF
\ENDFOR
\end{algorithmic}
\end{algorithm}

The proximity request sub-protocol
is divided into three steps, whose pseudo-code is illustrated in Protocol~\ref{prot:hnc}.
In Step $(i)$, user $A$ computes,
for each buddy $B$,
the set $S'$ of indexes of granules of $G_B$
such that, if $B$ is located in one of these granules,
then $B$ is in proximity.
More formally, $A$ computes the set of indexes $i$
such that the minimum distance $minDist$ between
the location of $A$ and $G_B(i)$ is less than or equal to $\delta_A$.
Then, in order to hide the cardinality of $S'$,
$A$ creates a new set $S$ by adding to $S'$ some non-valid 
randomly chosen 
indexes
(e.g., negative numbers).
This is done to increase the cardinality of $S$
without affecting the result of the computation.
The cardinality of $S$ is increased so that it is as large as the
number $sMax(G_B, \delta_A)$ that
represents the maximum number of granules of $G_B$ that intersect
with any circle with radius $\delta_A$.
Note that $sMax(G_B, \delta_A)$ can be computed
off-line since its values depend only on $G_B$ and $\delta_A$.
In the following, when no confusion arises, we use $sMax$ as a short notation for $sMax(G_B, \delta_A)$.

In Line~8, each element of $S$
is first hashed using the key $K^{ui}$,
which is obtained as the $ui$-th
value generated by the keystream initialized with $K_{B}$.
In this case $ui$ is the index of the update interval preceding
the current one.
Then, the result is encrypted,
using the commutative encryption function $C$ and
key $K_1$ that is randomly generated.
The element composed by the set $ES$ computed in Line~8,
$B$, and $ui$ is then added
to the set \emph{proxReq}.

Once the operations in Lines~4 to 9
are executed for each buddy $B$,
the set \emph{proxReq} is sent to the SP.

Upon receiving \emph{proxReq}, the SP
starts Step (ii).
For each tuple $\langle B, ui, ES\rangle$
in \emph{proxReq}, the SP
encrypts with the $C$ function each element of $ES$ using key $K_2$,
which is randomly generated.
The result is the set $ES'$.
Then, it retrieves the tuple $\langle B, ui, h_B \rangle$
updated by $B$ at the update interval $ui$.
In this tuple, $h_B$ is the value of the index
of the granule of $G_B$ where $B$ is located,
hashed with the key $K^{ui}$.
Since $ui$ is the update interval preceding the current one,
our location update policy assures that
a location update with update interval $ui$
has already been issued by every buddy $B$.
Finally, 
the SP encrypts $h_B$ with the commutative encryption function
$C$ using key $K_2$.
The resulting value $h'$ is added,
together with $B$ and $ES'$, to the set $proxResp$.

Once the computations at Lines~4 to 7
are executed for each buddy $B$,
the set \emph{proxResp} is sent to $A$.

In Step (iii), given the message \emph{proxResp}
received from the SP, $A$ computes the proximity of her buddies.
For each tuple $\langle B, ES', h' \rangle$,
$A$ obtains $h''$ as the encryption of $h'$ with $C$ and the key $K_1$
and checks if the result is in $ES'$.
If this is the case, then $B$ is in proximity,
otherwise he is not.

More formally,  $h'' \in ES'$
if and only if the granule of $G_B$ with index $i$
containing $B$ is in $S'$, that is equivalent to $B$ being in proximity.
Indeed, for each buddy $B$,
we recall that:
\[h'' = C_{K_1}(C_{K_2}(h_B))\]
and
\[ES' = \bigcup_{i \in S}(C_{K_2}(C_{K_1}(H_{K^{ui}}(i)))\]
Consequently, due to the commutative property of the encryption function,
$h'' \in ES'$ if and only if
\[h_B \in \bigcup_{i \in S} H_{K^{ui}}(i)\]
Since $h_B$ and the elements of the set are hashed using the same key $K^{ui}$,
$h_B$ is in the set if and only if $i \in S$.
Since $S=S' \cup S''$ and $i \not \in S''$
(because $S''$ contains invalid integers only while $i$ is a valid integer)
then $i \in S$ if and only if $i \in S'$.
By definition of $S'$, this implies that $B$ is in proximity.

\begin{figure}[htb]%
\centering
\includegraphics[scale=0.5]{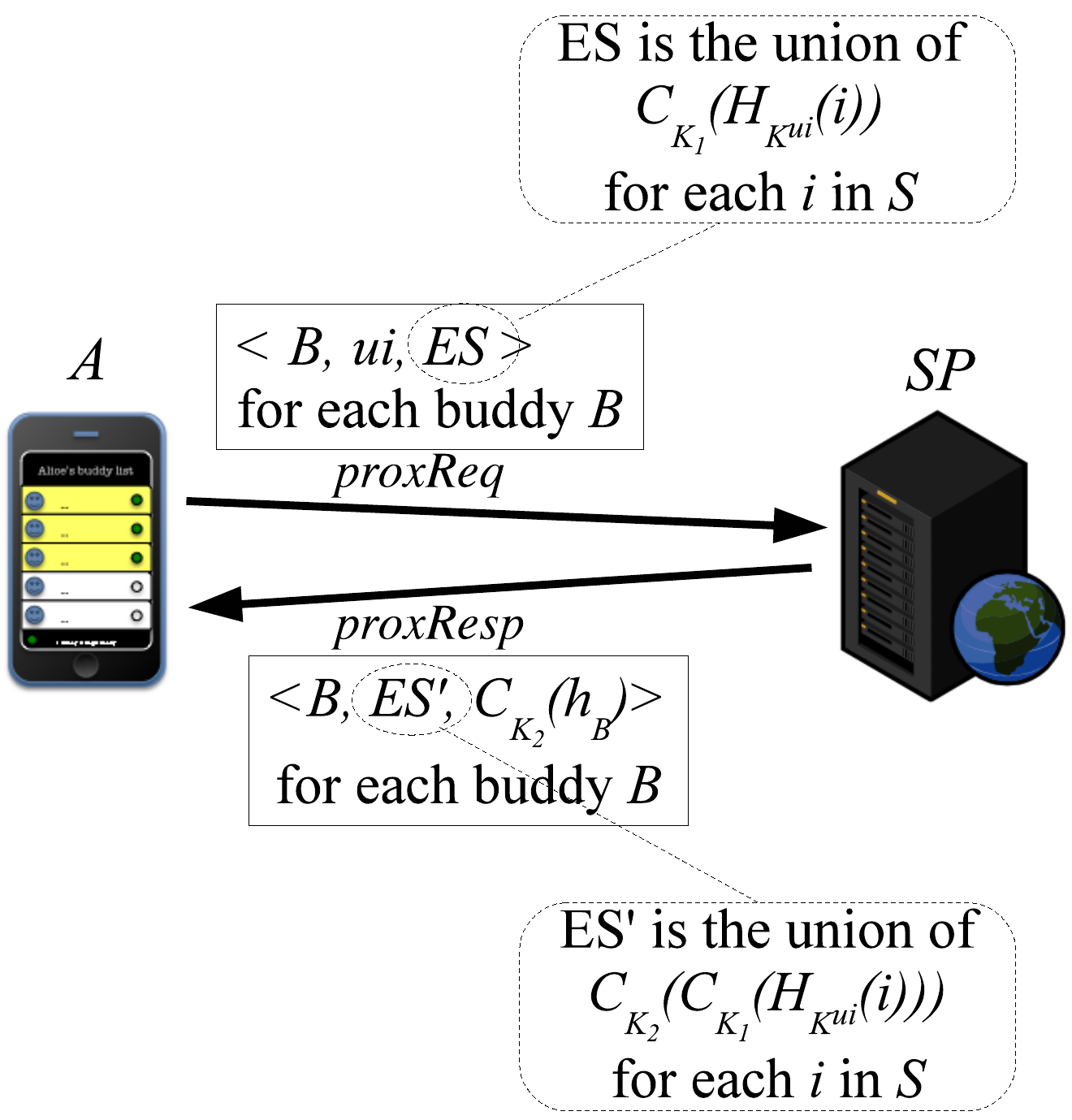}
\caption{Proximity request sub-protocol in \chnh.}
\label{fig:CHideAndHash}
\end{figure}

Figure~\ref{fig:CHideAndHash} shows the messages exchanged during the
proximity request sub-protocol of \chnh.

\subsection{Contrasting velocity attacks and other background knowledge}
\label{sub:velocity}
It is easily seen that our location update policy,
based on fixed length update intervals,
makes the probability that a location update is issued
independent from the location from where it is issued.
This is an important property used in Section~\ref{sec:analysis}, together with others, to prove
the safety of our solutions under the adversary models we consider.

Clearly, if the adversary had arbitrary background knowledge, there
would not be any technique that could guarantee privacy. However, it
is interesting to consider some other forms of knowledge that the
adversary could use.  
With respect to previous proposals, our defenses are resistant to an important type of background knowledge: a-priori distribution of the users' locations. 
There are, however, other types of knowledge that may be interesting to consider as, for example, the \emph{time-dependent} a-priori
location knowledge. This includes knowledge on the \emph{relative}
position of users at a certain time, as well as a-priori probability of user
movements. With this kind of knowledge it is also possible to perform attacks based on the velocity of users.
Consider Example~\ref{ex:velocity}.

\begin{example}
\label{ex:velocity}
User $A$ sends two location updates
in two consecutive update intervals $i$ and $j$
from granule $g_1$ and $g_2$, respectively.
Her buddy $B$ issues a proximity request
in each update interval and
discovers the granule where $A$ is located.
So far, no privacy violation occurred for $A$.
However, if $B$ knows
that $A$ moves at most with velocity $v$,
then he can exclude that $A$
is located in some locations $l$ of $g_2$.
Indeed, $B$ knows that the temporal
distance between the two location updates of $A$
is equal to the length $T$ of the update period.
Now $B$ can exclude that $A$ is located in any
location $l$ of $g_2$ such that
the time required to move from any point 
of $g_1$ to $l$ with velocity $v$
is larger than $T$.
Hence $B$ violates the privacy requirement of $A$.
\end{example}

The problem in Example~\ref{ex:velocity} arises when the adversary
knows the maximum velocity of a user.
Velocity-based attacks have been recently considered independently from proximity services \cite{GhinitaGIS09}, but the application of those solutions in our framework would lead to the release of some location information to the SP.
In the following we show how to adapt our location update policy
to provide protection preserving our privacy properties in the specific case
in which the adversary knows the maximum velocity $v$ of a user.
  
Let $tMax(g_1, g_2)$ be the maximum time required to move
at velocity $v$
from each point of granule $g_1$ to each point of granule $g_2$.
The problem of Example~\ref{ex:velocity} arises when the temporal distance between
two location updates issued from two different granules $g_1$ and $g_2$
is less then $tMax(g_1, g_2)$.
The problem can be solved by
imposing that $A$, after entering $g_2$,
randomly reports $g_1$ or $g_2$
as the granule where she is located
until time $tMax(g_1, g_2)$ elapses
from the last location update in $g_1$.
This solution is a form of temporal generalization
as it adds uncertainty to the adversary,
about when the user crosses the border between
$g_1$ and $g_2$.
More specifically, the adversary is unable
to identify the exact instant in which the user
crossed the border 
in a time interval of length at least $tMax(g_1, g_2)$.
Consequently, by definition of $tMax(g_1, g_2)$,
the adversary cannot exclude that $A$ moved
from any point of $g_1$ to any point of $g_2$.

The extension of our defense techniques to other forms of 
background knowledge is one of the subjects for future work.

\section{Analysis of the protocols}
\label{sec:analysis}

The main goal of our techniques is to guarantee
the satisfaction of users' 
privacy requirements under the
given adversary models.
In Section \ref{sub:privacy} we prove that our two protocols have
this property.

However, there are other important parameters to be considered in an
evaluation and comparison among protocols that satisfy the privacy requirements.
In general, the higher the privacy provided by the protocol, the
better is for the users; since location privacy in our model 
is captured by the {\em size of the uncertainty region}, in Section
\ref{sub:size} we consider this parameter.

A second parameter to be considered is {\em service precision}.
The percentage of false positives and false negatives introduced by a
specific protocol must be evaluated. This is considered in Subsection \ref{sub:qos}.

Last but not least, it is important to evaluate the overall {\em system cost}, including
computation and communication, with a particular attention to
client-side costs. This is considered in Subsection \ref{sub:costs}.

The proofs of the formal results presented in this section are in Appendix~\ref{sec:proofs}.
\subsection{Privacy}
\label{sub:privacy}

We first analyze the privacy provided by \Chns and \Chnc in Section~\ref{sub:no-collusion} considering the adversary models presented in Section~\ref{sec:prob} 
under the \emph{no-collusion} assumption, i.e., assuming that the SP does not collude with
the buddies and that the buddies do not collude among themselves.
Then, in Section~\ref{sub:collusion} we show the privacy guarantees
provided by the two algorithms in the more general case of possibly colluding adversaries.

\subsubsection{Satisfaction of privacy requirements} 
\label{sub:no-collusion}
We first analyze the \Chns protocol.
Since the private key $K_A$ is only known to $A$ and to the buddies of $A$, the SP is not able to decrypt the index of the granule where $A$ is located. Analogously, the SP is not able to obtain location information about $A$'s buddies and, in particular, does not obtain any information about the distance between $A$ and her buddies.

We now state a formal property of the \chns that is used in the formal proof of the above observations.

\begin{lemma}
\label{le:location_message_independence}
The \chns protocol ensures that under any a-priori knowledge $pri_{A}$, the following two random variables are probabilistically independent: (1) The binary random variable $ur(A)$: an update/request is sent by user $A$, 
and (2) random variable $loc_A$, i.e., the location of $A$, of any distribution. Formally, we have
$$P(ur(A)|loc_A, pri_{A}) =P(ur(A)|pri_{A}),$$
for any a-priori location knowledge $pri_{A}$ and location random variable $loc_A$ for user $A$.
\end{lemma}

Note that we are assuming discrete time and discrete location. A continuous case can be formalized and proved equally easily. 
Also, this lemma does not concern the type or content of a message sent by $A$, but just the fact that a message is sent by $A$.

Another property we use to prove our safety result is provided by the encryption algorithms, via the information theoretical notion of ``perfect secrecy'' \cite{bishopbook03:perfect_secrecy}. Intuitively, perfect secrecy for an encryption algorithm means that given ciphertext $c$, each plaintext $p$ has the same probability to be encrypted to $c$ (posterior), with a randomly chosen key, as the probability of $p$ to be used in the first place (prior). That is, $P(p|c)=P(p)$.  Equivalently, given plaintext $p$, each ciphertext $c$ has the same probability to be the encryption of $p$ (posterior), with a randomly chosen key, as the probability of $c$ to appear in the first place as ciphertext (prior). That is, $P(c|p)=P(c)$. Applied to our situation, when SP receives a message $\langle A, ui, E_{K^{ui}}(l)\rangle$, since $K^{ui}$ is hidden from the SP and can be chosen arbitrarily, the probability that SP receives any other message of the form $\langle A, ui, E_{K^{ui}}(l')\rangle$ is the same. 

Most of practical encryption algorithms do not have the theoretical perfect secrecy, but use computational hardness to achieve secrecy in the sense that it is computationally very hard (or impractical) to derive the plaintext from the ciphertext. 
Intuitively, 
$P(p|c)=P(p)$ holds because $c$ does not yield any information about $p$. Therefore, we use the simplifying, practical assumption that the encryption methods we use do give us perfect secrecy. 

The above perfect secrecy discussion applies to single messages. When dealing with multiple messages, correlation between plaintexts may reveal secrets when the same key is used. This is the classical scenario of repeated key use problem, and one solution to this problem is to use so-called one-use-pad or keystreams as we do in our proposed protocols. As each key is only used once, encrypted messages are independent to each other when perfect secrecy is assumed.

From
the above discussion and assumptions, Lemma~\ref{le:message_independence} follows. 
Since the lemma involves random variables on messages, we need to specify the {\em message space} for these variables. 
We consider the randomness of the messages to be on the encrypted part, while other parts are fixed. Formally,
we call each sequence $\langle B_1, ui_1\rangle, \ldots, \langle B_n, ui_n\rangle$,
where $B_j$ is a user and $ui_j$ is a time interval, a {\em (message set) type}. (Recall that a message is of the form $\langle B, ui, ES\rangle$.)
The messages of the same type differ on the encrypted part of the messages and constitute a message space.
When a generic message $M$ is mentioned, we assume it is a variable over all the messages with a specific type. 

\begin{lemma}
\label{le:message_independence}
Given messages $M=M_1\cup M_2$ issued in the \chns protocol, where $M_1\cap M_2=\emptyset$,  we have $$P(M|loc_A, pri_{A})=P(M_1|loc_A, pri_{A})*P(M_2|loc_A, pri_{A}),$$
for all a-priori knowledge $pri_{A}$ and location $loc_A$ for user $A$.
\end{lemma}

With Lemma~\ref{le:location_message_independence}, perfect secrecy, and Lemma~\ref{le:message_independence}, we now show a main result, namely, 
the SP does not acquire any location information as a consequence of a location update or a proximity request using the \chns protocol.
The following formal results implicitly refer to our adversary models that, in particular, assume that the SP has no background knowledge other than the protocol, the a-priori distribution, and the granularities.

\begin{theorem}
\label{th:hnsSP}
Let $A$ be a user issuing a sequence of location updates and proximity
requests following the \chns protocol.
Then, $A$'s privacy requirement is satisfied with respect to the SP.
\end{theorem}

We now turn to the location information acquired by the buddies. 
In the \chns protocol, a user $A$ issuing a proximity request does not send any location information, hence her buddies, even if malicious, cannot violate her privacy requirements. When the same user runs the location update subprotocol in \chns, her buddies can only obtain the granule at the granularity $G_A$ in which $A$ is located. As a consequence, the privacy
requirement of $A$ is guaranteed. This is formally stated in
Theorem~\ref{th:hnsbuddies}.

\begin{theorem}
\label{th:hnsbuddies}
Let $A$ be a user issuing a sequence of location updates and proximity
requests following the \chns protocol.
Then, $A$'s privacy requirement is satisfied with respect to each of $A$'s buddies.
\end{theorem}

We consider now the \chnc protocol.
Since $K_{A}$ is only known to $A$ and her buddies, the SP is not able to acquire the location information provided by $A$ during a location update. This follows from Theorem~\ref{th:hnsSP}. The difference of the \chnc from the \chns is that when $A$ issues a proximity request in \chnc, an encrypted message is sent to the SP. However, due to the property of the secure computation protocol in \chnc, the only information that the SP acquires about the set provided by $A$ is its cardinality. Actually, the cardinality of this set is always $S_{MAX}$ that, by definition, depends only on $\delta_A$ and $G_B$, and not on the actual location of $A$ or $B$. Consequently, the SP does not acquire any information about the location of $A$ and $B$, including their distance. Theorem~\ref{th:hncSP} formally states this property.

\begin{theorem}
\label{th:hncSP}
Let $A$ be a user issuing a sequence of location updates and proximity
requests following the \chnc protocol.
Then $A$'s privacy requirement is satisfied with respect to the SP.
\end{theorem}

Similarly to the \chns protocol, in \chnc each buddy of $A$ can only obtain location information derived from $A$'s location update.
It is worth noting that in the \chns protocol, each time $B$ issues a proximity request, he obtains the granule of $G_{A}$ where his buddy $A$ is located. Differently, using the \chnc protocol, $B$ only gets to know whether the granule where $A$ is located is one of those in $S_{A}$. This means that, if $A$ is not in proximity, then $B$ only learns that $A$ is not in any of the granules of $S_{A}$. Otherwise, if $A$ is in proximity, $B$ learns that $A$ is in one of the granules of $S_A$, without knowing exactly in which granule she is located.
This is formally stated in Theorem~\ref{th:hncbuddies}.

\begin{theorem}
\label{th:hncbuddies}
Let $A$ be a user issuing a sequence of location updates and proximity
requests following the \chnc protocol.
Then, $A$'s privacy requirement is satisfied with respect to each of $A$'s buddies.
\end{theorem}

 In Section~\ref{sec:exres} we show that, on average, \chnc provides more privacy with respect to the buddies than \chns, but at extra costs, making each protocol more adequate than the other based on user preferences and deployment modalities. 

\subsubsection{Privacy in case of possibly colluding adversaries}
\label{sub:collusion}
We now consider the case in which our reference adversaries can collude,
and we analyze the privacy guarantees of the \Chnc and \Chns protocols in this scenario.

First, consider the case in which two buddies $B$ and $C$ collude 
to violate the privacy of a user $A$.
The problem can be easily extended to consider more buddies.
Let $l_B$ be the set of possible locations of $A$
obtained by $B$ as a result of a proximity request.
Let $l_C$ be the analogous information acquired by $C$ during the same update interval.
Since $B$ and $C$ collude, they can derive that $A$
is located in $l_B \cap l_C$.
However, due to Theorem~\ref{th:hncbuddies},
given $G_A(i)$ the granule where $A$ is located,
it holds that
$l_B \supseteq G_A(i)$ and  
$l_C \supseteq G_A(i)$ 
(recall that $G_A$ is the privacy requirement of $A$ with respect to the buddies).
Consequently,  $l_B \cap l_C \supseteq G_A(i)$ and hence
the privacy requirement of $A$
is guaranteed also in the case $B$ and $C$ collude.

Now, consider the case in which the SP colludes with one or more buddies.
For example, if one of the buddies shares the secret key $K_A$ with the SP, the SP can learn the granule where $A$ is located.
In this case, the 
privacy requirement of $A$ with respect to the SP
is not guaranteed.
Nevertheless, even if the SP knows $K_A$, he cannot discover the location of $A$
within the granule of $G_A(i)$ where $A$ is located.
This is because, by the definition of the two protocols,
every message issued by $A$ does not depend on the location of $A$ within $G_A(i)$.
Consequently, the 
privacy requirement with respect to the buddies is still guaranteed.
This means that the lowest privacy requirement
of the two colluding entities is preserved
and this is the best that can be achieved in case of collusion.

\subsection{Service precision}
\label{sub:qos}

The techniques proposed in the literature as well as the
techniques we propose in this paper, generalize the
location of one of the two users to an area.
When proximity is computed,
the exact location of that user within the area is not known.
Hence, proximity is evaluated as the distance between
a point and a region\footnote{In previous work,
the location of both users is generalized
and proximity is computed between two regions.}.

\begin{figure}[ht]
\centering
\subfigure[$B$ is in proximity of $A$]{
\label{fig:inProx}
	\includegraphics[scale=0.25]{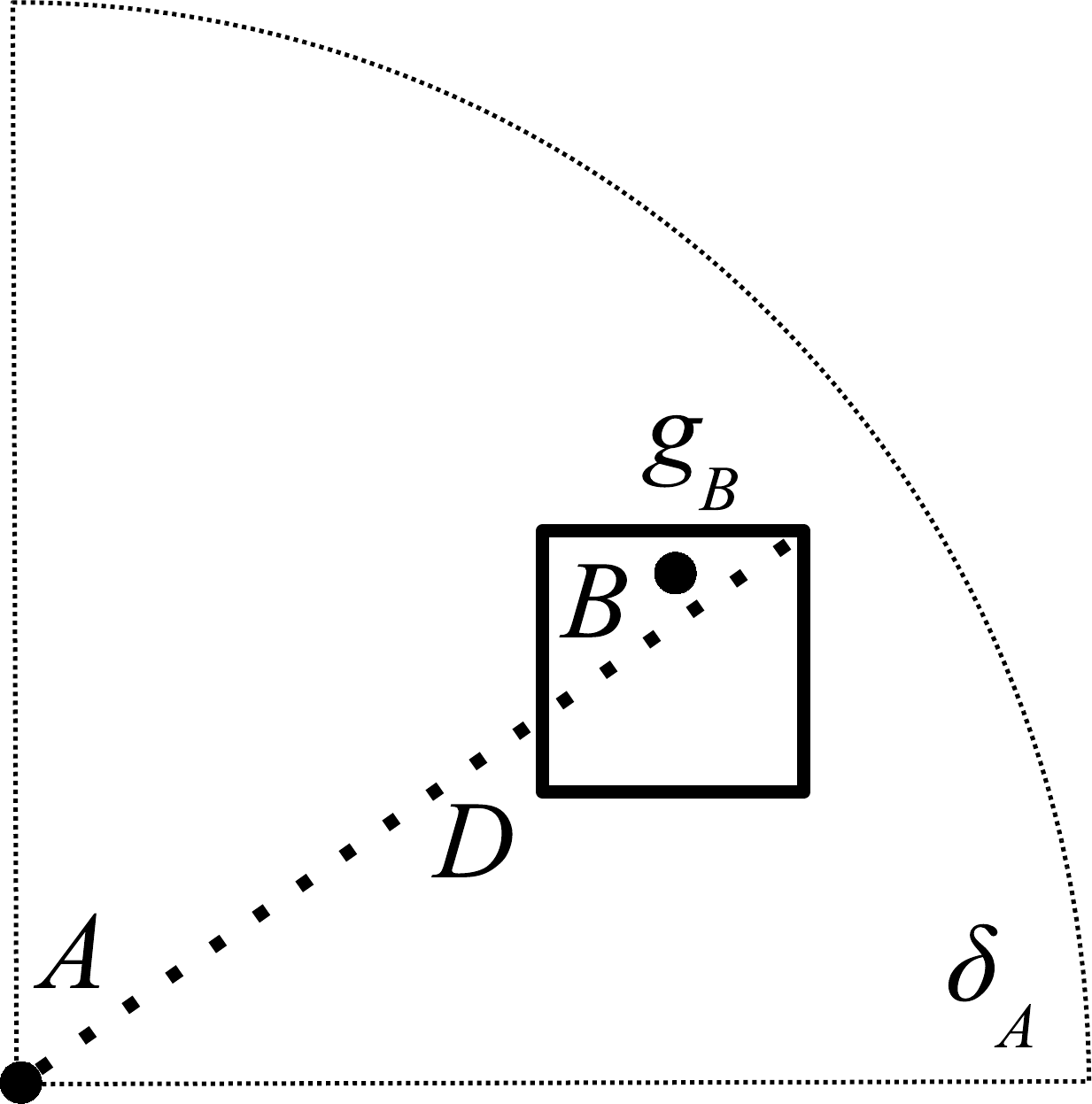} %
}
\subfigure[$B$ is not in proximity of $A$]{
\label{fig:notInProx}
	\includegraphics[scale=0.21]{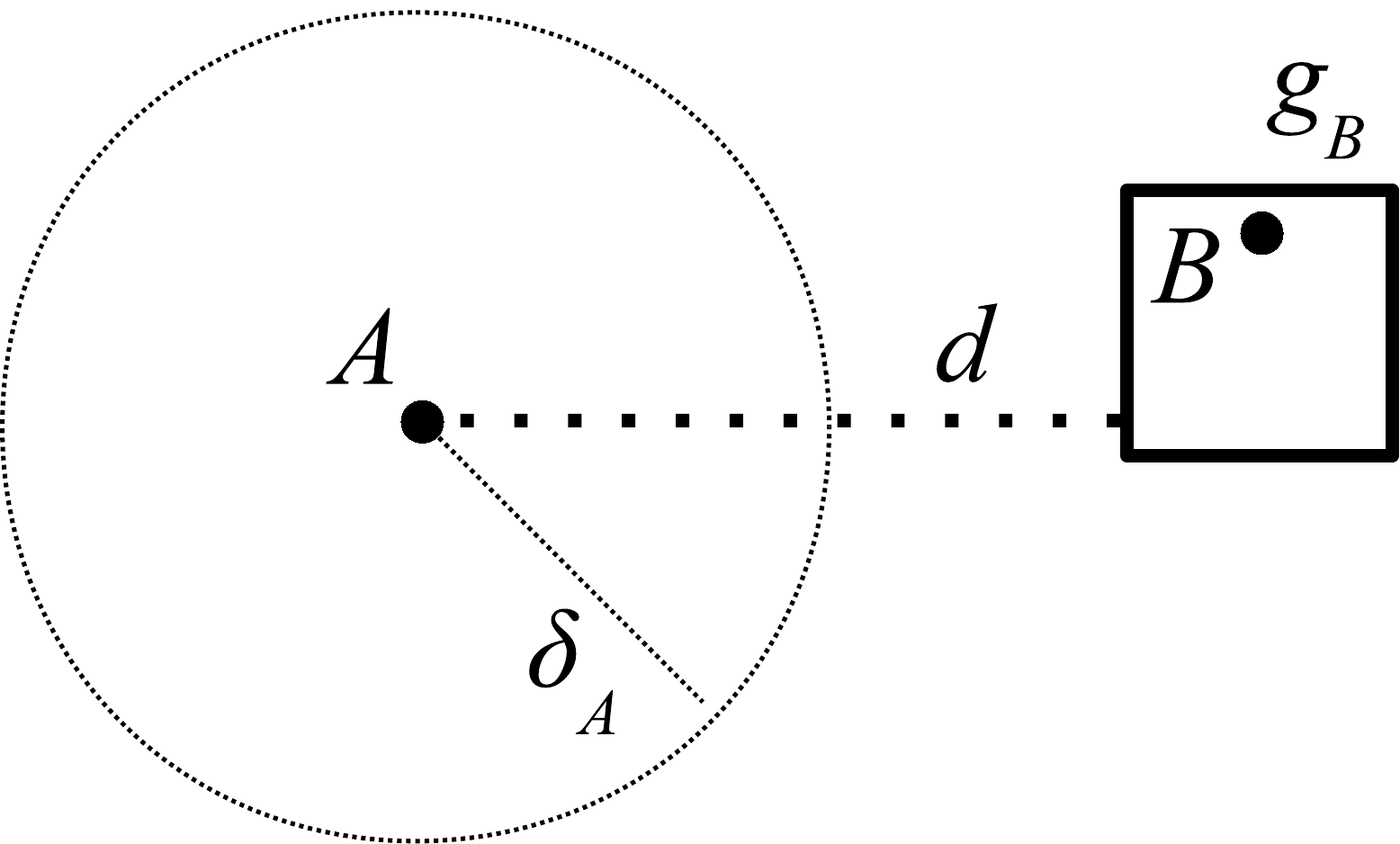} %
}
\subfigure[$B$ is \emph{possibly} in proximity of $A$]{
\label{fig:SPdoesntKnow}
	\includegraphics[scale=0.30]{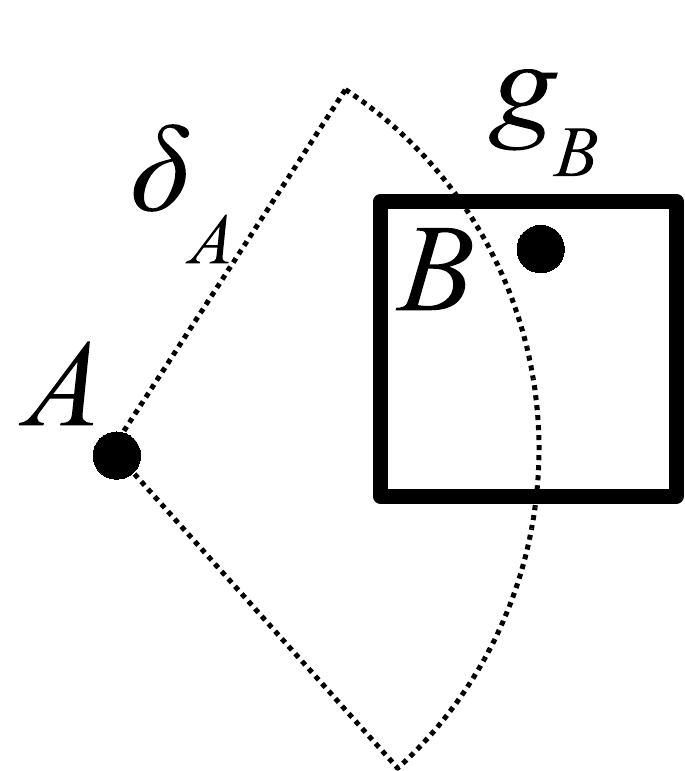} %
}
\caption{Regions $L_A$ and $L_B$}%
\label{fig:deltaExample}
\end{figure}

Consider how it is possible to compute the proximity
between a user $A$ whose exact location is known
and a user $B$ whose location is only known to be in region.
It is easily seen that
if the maximum distance between the point and the region
is less than the proximity threshold,
then the two users are in proximity,
independently from where $B$ is located 
within the region.
Figure~\ref{fig:inProx} shows an example of this situation.
On the contrary, if the minimum distance
is larger than the distance threshold,
then the two users are not in proximity.
Figure~\ref{fig:notInProx} graphically shows that this happens
when no point of the region containing $B$
is in proximity of $A$.
If none of the two cases above happen
(i.e., the threshold distance is larger than
the minimum distance and less than the maximum distance),
we are in presence of an \emph{uncertainty case},
in which it is not possible to compute
whether the two users are in proximity
without introducing some approximation in the result.
For example, Figure~\ref{fig:SPdoesntKnow} shows that
if $B$ is located close to the bottom left corner of the region
then $B$ is in the proximity of $A$, otherwise he is not.

The choice we made in the presentation of our protocols is 
to consider two users as in proximity 
in the \emph{uncertainty case}.
The rational is that in this case it is not possible to exclude that
the users are not in proximity.
Previous approaches (\cite{ZhongPET07,MascettiMDM09}) facing a similar
issue have adopted the same semantics.

One drawback of this \mindist semantics
is that it generates false positive results
and this may be undesirable in some applications.
Indeed, if user $B$ is reported to be in proximity of $A$,
then $A$ may decide to contact $B$ (e.g., through IM).
This may be annoying for $B$, if he is not actually in proximity.
Consider, for example, the case in which the location of $B$
is reported at the granularity of a city: $B$ is always reported as
in proximity of $A$ when $A$ is in the same city, independently
from the proximity threshold chosen by $A$.

\begin{figure}[htb]
\centering
\includegraphics[scale=0.55]{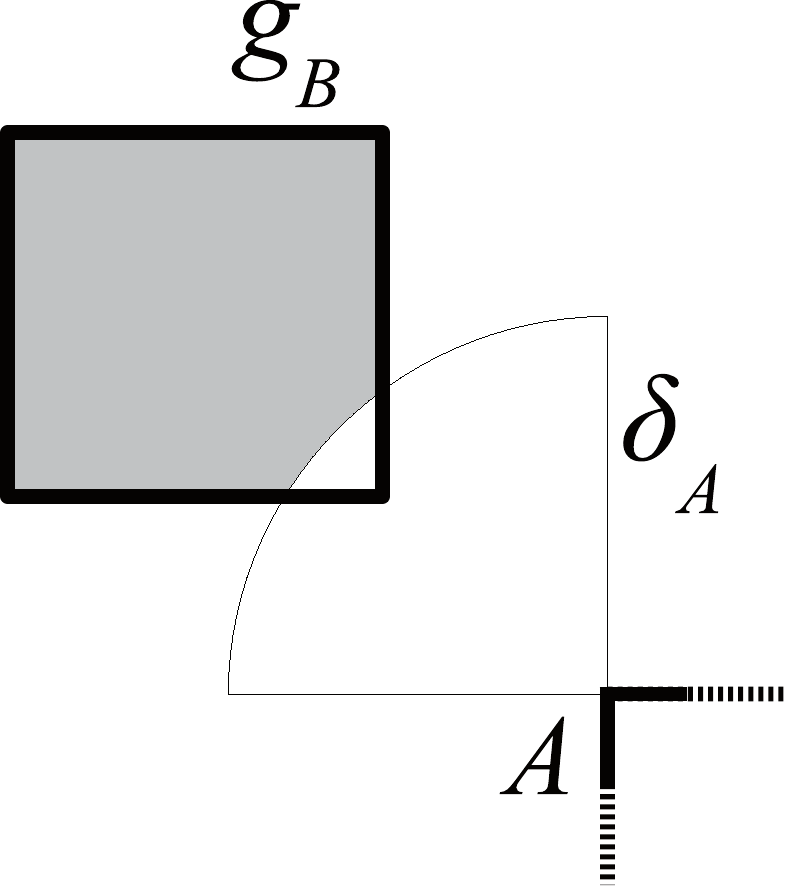}
\caption{Approximation incurring with the \mindist semantics}
\label{fig:minDistApp}
\end{figure}

An alternative semantics, that we name \maxdist semantics, solves this problem.
The idea is to consider two users as in proximity
only when it is certain that they are actually in proximity.
This happens when the maximum distance between their areas
is less than the distance threshold.
While this approach does not generate any false-positive,
it does produce false-negatives.
The two semantics above have a common drawback:
in certain cases it happens that the probability
of providing a false result is larger than
the probability of providing a correct result.
Consider the example depicted in Figure~\ref{fig:minDistApp}
in which the \mindist semantics is considered.
User $B$ is considered in proximity but
the answer is wrong if $B$
is located in the region colored in gray.
Assuming a uniform distribution of $B$ inside $g_B$,
it is much more likely to have an incorrect result,
rather than a correct one.
An analogous problem can arise for the \maxdist approach.

The percentage of false results can be minimized
by considering user $B$ as in proximity
only when at least one half of the area
is actually in proximity.
The drawback of this \mostprox
semantics is that it incurs in both false positive and false negative results.

Our protocols are designed so that it is very easy to change the
current proximity semantics. Since this can be done client-side,
without the need for changes server-side nor in the code other peers
are running, the semantics can be potentially chosen through the user
interface at any time.

We analytically measured the impact of
the different semantics on the accuracy of our protocols 
by calculating the \emph{expected precision} and the \emph{expected recall}. 
The expected precision is defined as the probability 
that a buddy reported to be in proximity according to a given semantic
is actually in proximity. Vice versa, the expected recall is defined as the 
 probability that a buddy actually in proximity is reported to be in 
proximity according to a given semantic.

\begin{figure}[htb]%
\centering
\scalebox{1}{\input{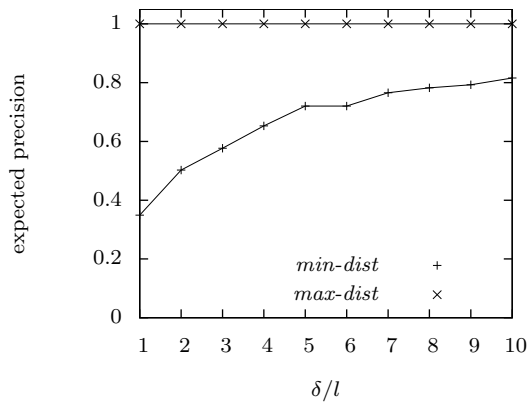}} %
\caption{Expected precision}
\label{fig:prec_edge}
\end{figure}

\begin{figure}[htb]%
\centering
\scalebox{1}{\input{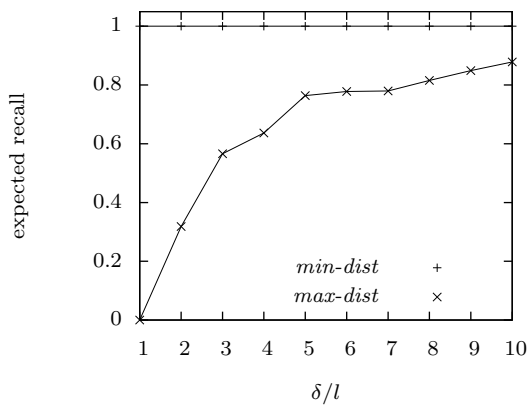}} %
\caption{Expected recall}
\label{fig:rec_edge}
\end{figure}

Figures~\ref{fig:prec_edge} and \ref{fig:rec_edge} show the minimum expected precision 
and recall for the \mindist and the \maxdist
semantics.
Both measures depend on the ratio between $\delta$ and the
area of the granules in which a user is considered in proximity.
For this analysis we considered a grid-like granularity containing 
cells having edge of size $l$ and we assume users are uniformly distributed.
As can be observed in Figure~\ref{fig:prec_edge}, the \maxdist
semantic has always precision equal to $1$. This is because all the buddies 
considered in proximity are always actually in proximity. The \mindist has precision of about $1/3$ when the values of $\delta$ and $l$ 
are equal, and this value grows logarithmically when $\delta$ is larger than 
$l$.
The analysis of expected recall (Figure~\ref{fig:rec_edge}) shows that the \mindist has always recall equal to $1$. This is because if a buddy is actually in proximity, it is always reported in proximity using this semantic. The \maxdist semantic, on the contrary, has a minimum expected recall equal to $0$ when $\delta$ and $l$ are equal. This is because, with this parameters, it can happen that no cells of size $l$ are fully contained in a circle having radius $\delta$. However, the recall of the \maxdist grows more rapidly than the precision of the \mindist.

\subsection{Size of uncertainty regions}
\label{sub:size}

As already discussed in Section~\ref{sub:privacy}, our protocols are
proven to always guarantee the 
privacy requirement with
respect to the buddies. However, the main difference between our two
protocols consists in the fact that \chnc can provide
additional privacy with respect to one buddy. For example, if a user
$A$ issues a proximity request using \chnc, and a buddy $B$
is reported as being not in proximity, $A$ only learns that $B$ is not
located in any of the granules considered in proximity (i.e., the ones
included in $S$). The resulting uncertainty region of $B$, in this case,
is equal to the entire space domain minus the region identified by
$S$. 
When $B$ is reported to be in proximity, $A$ learns that $B$
is located in one of the granules of $S$, but not exactly in which of
those granules. Therefore, the uncertainty region in this case is
given by the region identified by $S$. The size of this region
depends on the value $\delta_A$, on the area of the granules in $G_B$, and on
the distance semantics chosen by $A$.
In order to show how the size of the uncertainty region is affected by
these parameters, we simplify the analysis by considering  grid-like
granularities, similarly to Section~\ref{sub:qos}.
Each granularity is a grid identified by the size $l$ of the edge of
its cells. 

\begin{figure}[htb]%
\centering
\scalebox{1}{\input{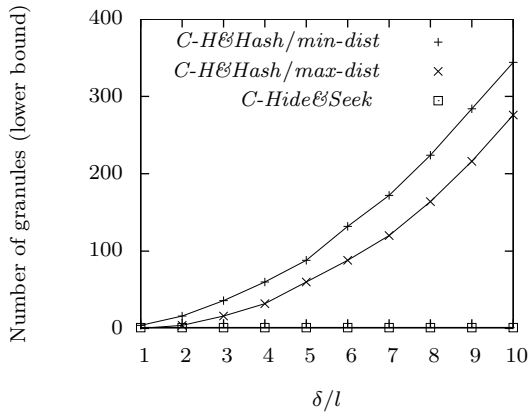}} %
\caption{Privacy with respect to a buddy}
\label{fig:privacy_delta}
\end{figure}

Figure~\ref{fig:privacy_delta} shows the 
additional privacy
achieved by \chnc for different values of $\delta / l$.
The additional privacy is measured as the lower bound of the number of granules in $S$.
As can be observed, using both semantics, the additional privacy grows
when $\delta$ is larger than $l$. This means, for example, that if 
$\delta$ is $5$ times larger than $l$, then the actual size of the
uncertainty region of $B$ is $60$ (or $88$) times larger than the
minimum privacy requirement if $A$ is using the
\maxdist (or \mindist, resp.) semantics.

\subsection{System costs}
\label{sub:costs}
We separately evaluate the computation and communication costs
involved in running the two proposed protocols. 
The analytical evaluation reported here is complemented with
experimental results in Section \ref{sec:exres}. 

\subsubsection{\Chns} 

In order to perform a location update, a user needs to compute the
index of the granule where she is located. The time complexity of this
operation depends on the data structure used to represent
granularities. As we shall show in Section~\ref{sec:exres}, with our
implementation of the granularities this operation can be performed in
constant time. The complexity of the encryption operation depends on
the encryption function and on the length of the encryption
key. Considering a fixed key length, the encryption of the index of
the granule can be performed in constant time.
Since the SP only needs to store the received information, the
expected computational complexity is constant.
The communication cost is constant and consists in an encrypted integer value.

For what concerns the cost of a proximity request on the client side,
for each buddy the issuing user needs to decrypt the index and to
compute the distance of the granule with that index from her
location. In our implementation these operations can be performed in
constant time and hence the time complexity of the proximity request
operation on the client side is linear in the number of buddies.
On the SP side, the computational cost to retrieve the last known
locations of the buddies is linear in the number of buddies.
The communication consists in one request message of constant size
from the user to the SP, and of one message from the SP to the user
with size linear in the number of buddies.

\subsubsection{\Chnc}

The cost of a location update operation on the client
is similar to the cost of the same 
operation using \chns, since the only difference is that a hashing 
function, which can be computed in constant time, is applied 
instead of the encryption function. Like in \chns, the SP only needs to store
the received information. 
Hence, computational costs of a location update are constant both for
the client and for the SP. The communication cost is constant, as the
only exchanged message consists in a hashed value.

On the client side, a proximity request from $A$ requires, for each buddy
$B$, the computation of the granules of $G_B$ which are considered in
proximity, the hashing, and the encryption of a number of granule
indexes in the order of $sMax(G_B,\delta_A)$. 
The value of $sMax$ can be pre-computed for a given
granularity. The computation of the granules considered in proximity
can be performed in constant time in our implementation, using grids
as granularities. 
The computation of the hashing and the encryption functions can also
be performed in constant time, hence the time complexity of a
proximity request is linear in the number of buddies times the maximum
among the $sMax$ values for the involved granularities. When the
client receives the response from the SP, the result computation
performed by $A$ for each buddy $B$ requires the encryption of a
number (the encrypted value sent by the SP), 
and the lookup of the encryption in a set of encrypted
values with cardinality $sMax(G_B, \delta_A)$. As the lookup in the set
of hashes requires at most $sMax$ operations, the time complexity is
then linear in the number of buddies times the maximum value of
$sMax$.
Hence, this is also the overall complexity on the client side.
On the SP side, the response to a proximity request from a user $A$
requires, for each buddy $B$, a) the retrieval and the encryption of the
hashed location of $B$, b) the encryption of the $sMax(G_B,\delta_A)$ hashed granule indexes sent
by $A$. As the encryption runs in constant time, the time
complexity is linear in the number of buddies times the maximum value
of $sMax$.

Regarding the communication costs, both of the messages involved in
the proximity request sub-protocol contain the encryption of a set of
a number of hashed values linear in the number of buddies times the
maximum value of $sMax$.

\section{System implementation}
\label{sec:implementation}

We implemented the techniques presented in Section~\ref{sec:tec} in a
system that provides proximity notification coupled with typical
instant messaging (IM) functionalities.  This implementation is the
evolution of the system developed for the \hnc protocol and it has
similar architecture, server and client applications
\cite{FreniSSTD09}.

\begin{figure}[ht]
\centering
\includegraphics[width=.9\columnwidth]{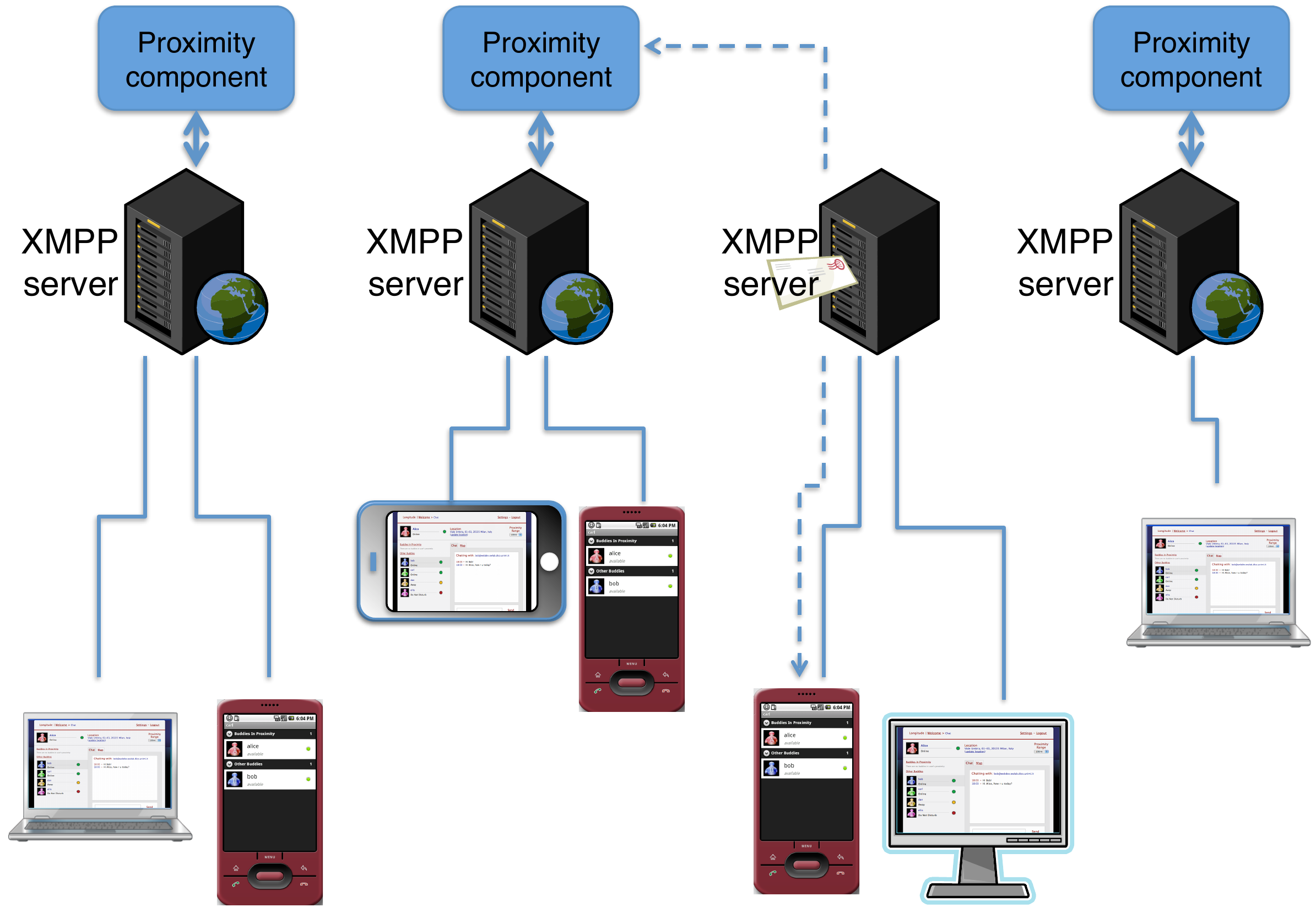} %
\caption{System architecture}%
\label{fig:xmpparch}
\end{figure}

The system is built as an extension of XMPP
(Extensible Messaging and Presence Protocol), 
an open standard protocol often used in commercial applications as a message oriented middleware \cite{rfc3920:xmpp}.
The system architecture is shown in Figure~\ref{fig:xmpparch}.
The choice of extending XMPP is driven by the following considerations.
First, the XMPP protocol can be easily extended to support
custom services and messages, like the proximity service, in our case.
In particular, by extending XMPP messages,
we designed a proper XML protocol for each of our technique.
In addition, the SP providing the proximity services
is implemented as a XMPP component
i.e., a pluggable entity that extends the default XMPP functionalities.
A second advantage is that the XMPP protocol already
includes standard sub-protocols for client-to-client communication and
for managing the list of buddies.
We used these sub-protocols as primitives in our implementation.
Finally, since the XMPP architecture is decentralized,
clients running on different servers can communicate with each other.
In our case, since a component acts as a special type of client,
this means that our proximity service is accessible to a
user registered to an existing XMPP service, including popular
IM services like Google Talk or Jabber.
This makes it possible to use,
in the proximity service, the same list of buddies
used in those IM services. Clearly, proximity can
be computed only for those buddies that are participating in
the same proximity service.

For what concerns the client, we developed a multi-platform web
application and an other application specifically designed for mobile
devices based on Apple iOS operating system, like iPhone or iPod touch.  In
addition to the typical functionalities of an IM application, the
clients implement the proximity protocols described in
Section~\ref{sec:tec} and provide the typical functionalities of a
full-fledged proximity service, including the detection of the client
user's location,
the notification of any buddies in proximity, and
the graphical visualization of the location uncertainty region for each buddy.

One of the issues emerged during the implementation of the \chnc
and \chns protocols concerns key management.  Indeed, both
protocols require that each user $A$ has a key $K_A$ that is shared
with all of her buddies, and it is kept secret to everybody else.  A
first problem is how $A$ can share her key with one buddy $B$ in a
secure manner.  This operation is required, for example, when the user
accesses the proximity service for the first time or a new buddy is
added to the buddy list.  To address this problem, we employ
standard public key cryptography techniques to encrypt, for each
buddy of a user $A$, the key $K_A$; After being encrypted, the key can be safely transmitted over an insecure channel.
The second problem is how to revoke a secret key.
For example, this is necessary when a buddy is removed from the buddy list,
or when the key is compromised.
In our implementation, 
in order to revoke a key, it is sufficient to generate a new secret key and
to send it to the authorized buddies.
The cost of sending a key to all the buddies
is clearly linear in the number of buddies.
In Section~\ref{sec:exres} we show that the costs to perform this
operation on a mobile device are sustainable.
In addition, it should be observed that the distribution of the key to all the buddies
is only needed when a user first subscribes to the proximity service
or when a buddy is removed from the buddy list.
These are very sporadic events during a typical IM service provisioning.

\section{Experimental results}
\label{sec:exres}

We conducted experiments to measure the performance of our protocols
and to compare them with the Pierre, \friendloc, \hns and \hnc
protocols \cite{ZhongPET07,SiksnysSSTD09,MascettiMDM09}.  We present
the experimental setting in Section~\ref{sub:exsetting}.  Then, in
Sections~\ref{sub:exqos}, \ref{sub:excosts} and \ref{sub:expriv} we
evaluate the protocols according to three evaluation criteria: quality
of service, privacy and system costs, respectively.

\subsection{The experimental setting}
\label{sub:exsetting}
The experimental evaluation of the protocols presented in this paper
was performed on a survey-driven synthetic dataset of user movements,
which was obtained using the \emph{MilanoByNight}
simulation\footnote{\url{http://everywarelab.dico.unimi.it/lbs-datasim}}.
We carefully tuned the simulator in order to reflect a typical
deployment scenario of a proximity service for geo-social networks:
$100,000$ potential users moving between their homes and one or more
entertainment places
in the city of Milan during a weekend night. The simulation also
models the time spent at the entertainment places, i.e., when no
movement occurs, following probability distributions extracted from
user surveys. All the test results shown in this section are obtained
as average values computed over $1,000$ users, each of them using the
service during the $4$ hours of the simulation.
Locations are sampled every $2$ minutes. 
The total size of the map is $215$ km$^2$ and the average density is $465$ users/km$^2$.
All the components of the system are implemented in Java.
Server-side test were performed on a 64-bit
Windows Server 2003 machine with $2.4$Ghz Intel Core 2 Quad processor
and $4$GB of shared RAM. 
Client-side tests were run on an iPhone 4 mobile device, running Apple
iOS 4.1 as operating system.  We implemented the symmetric encryption
and the hashing functions using the RC4 and MD5 algorithms,
respectively, while the RSA public key encryption algorithm was used
for the key distribution.

In the experiments we used grid-based granularities. Each granularity
is identified by the size of the edge of one cell of the grid. The
location-to-granule conversion operations required by our protocol can
be performed in constant time.
For the sake of simplicity, in our tests we assume that all the users
share the same parameters and that each user stays on-line during the
entire simulation.  Table~\ref{tab:parameters} shows the parameters
used in our experiments.  Note that the ``number of buddies''
parameter refers to the number of \emph{on-line} buddies that, for the
considered type of application, is usually significantly smaller than
the total number of buddies.
\begin{table}[ht]
\caption{Parameter values}
\label{tab:parameters}
\begin{center}
\small{
\begin{tabular}{|c|c|}
\hline
\textbf{Parameter} & \textbf{Values} \\
\hline
$\delta$ & 200m, \textbf{400m}, 800m, 1600m \\
\hline
Edge of a cell of $G$ & 100m, \textbf{200m}, 400m, 800m \\
\hline
Number of buddies & 10, 25, \textbf{50}, 75, 100  \\ 
\hline
\end{tabular}
}
\end{center}
\end{table}
\vspace{-1.3cm}
\subsection{Evaluation of the quality of service}
\label{sub:exqos}
\begin{figure*}[htb]
\centering
\subfigure[Pierre / \friendloc]{
\label{fig:pierre}
\includegraphics[scale=0.55]{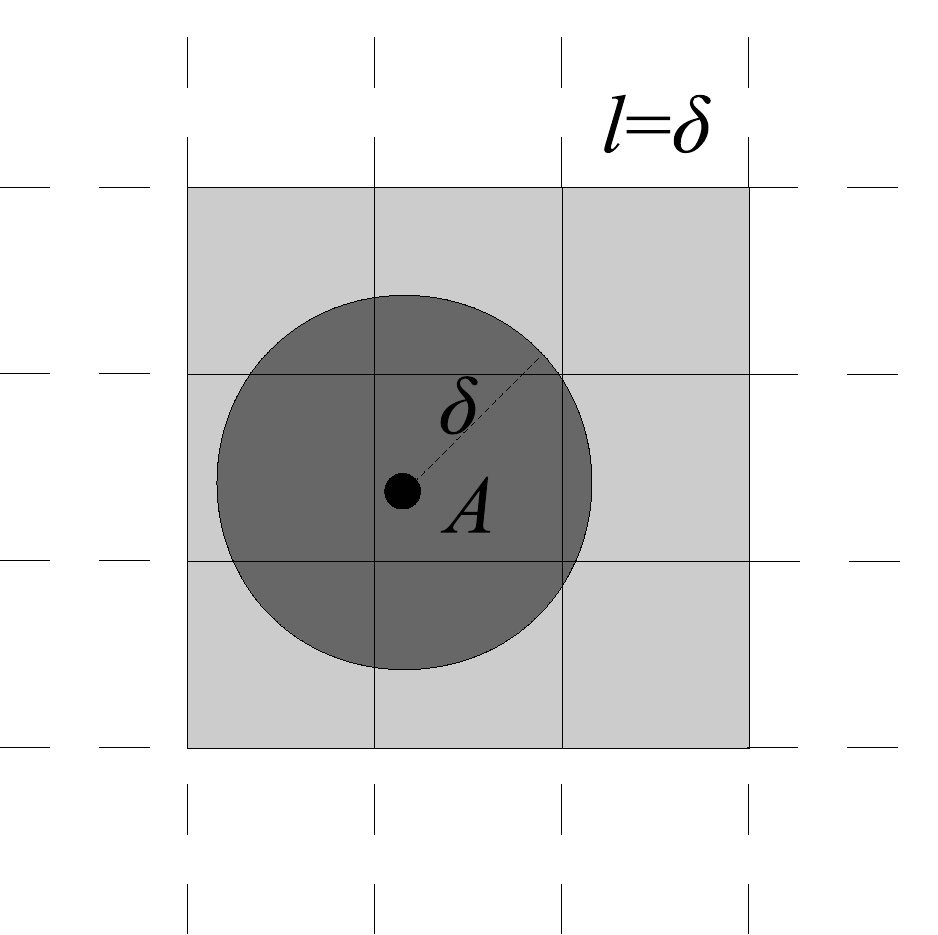}}
\hspace*{20pt}
\subfigure[\chns / \chnc]{
\label{fig:hncapprox}
\includegraphics[scale=0.55]{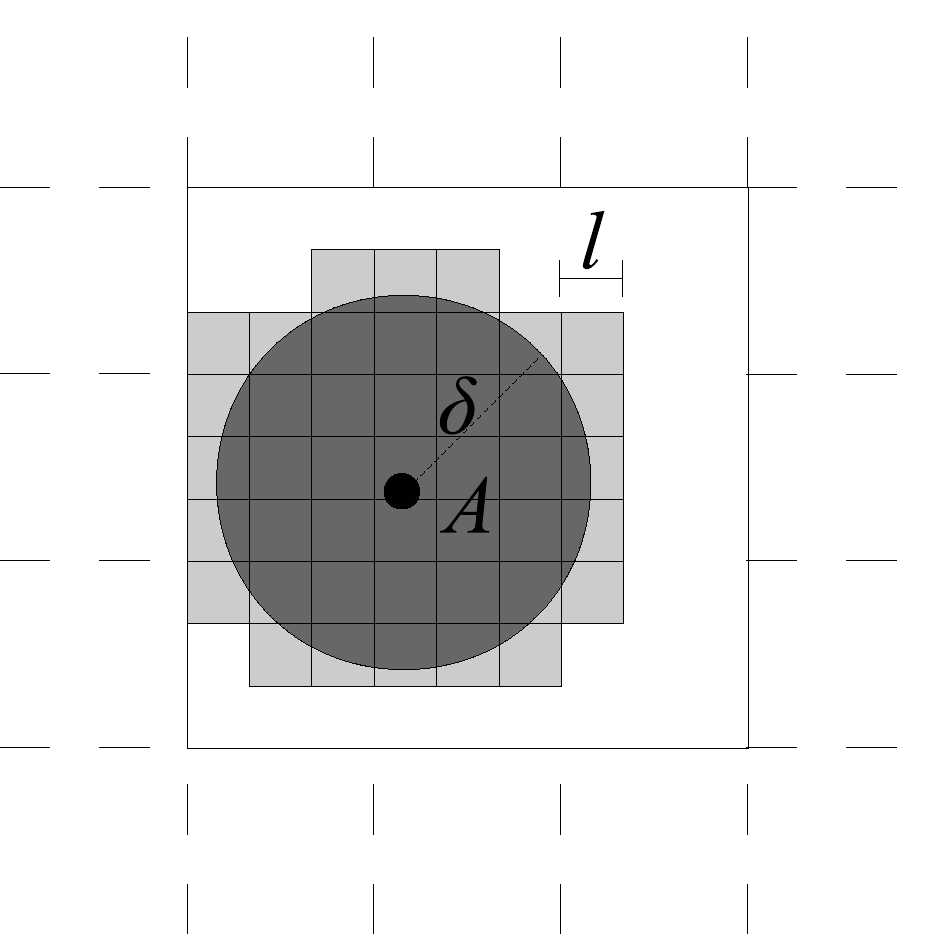}}
\caption{Examples of the \emph{granularity approximation}}%
\label{fig:deltaApprox}
\end{figure*}
The first set of experiments evaluate the impact of the techniques on
the \emph{quality of service}, by measuring the
exactness
of the answers returned by each protocol.  Indeed, two forms of
approximation are introduced by our protocols.  The \emph{granularity
  approximation} is caused by the fact that, when computing the
proximity between two users, the location of one of them is always
generalized to the corresponding granule of her privacy requirement
granularity.  The other approximation, which we call the
\emph{time-dependent approximation}, is due to the fact that, when a
user issues a proximity request with \chns, proximity is computed with
respect to the last reported location of each buddy.  The
approximation is introduced because the buddies have possibly moved
since their last location update.  Similarly, during the computation
of a proximity request with \chnc, the location transmitted by each
buddy during the previous update interval is used.

For what concerns the \emph{granularity approximation}, a similar
problem occurs with the Pierre and \friendloc protocols too.  Indeed,
both protocols, in order to detect proximity between buddies,
partition the domain space into a grid, with each cell having edge $l$
equal to the distance threshold $\delta$, that must be shared by the
users.  Then, a buddy $B$ is considered in proximity of $A$ whether
$B$ is located in the same cell as $A$ or in one of the $8$ adjacent
cells.  The approximation introduced by these techniques depends
entirely on the chosen value of $\delta$. Differently, in our
solutions, each user can choose her privacy requirements independently
from the value of $\delta$. For example, consider
Figure~\ref{fig:deltaApprox}. The black dot is the actual location of
user $A$.  The dark gray circle with radius $\delta$ is the area where
the buddies of $A$ are actually in proximity of $A$. The light gray
area is the region in which buddies are erroneously reported to be in
proximity\footnote{Here and in the following, we assume users of our
  protocols are choosing the \emph{minimum-distance}
  semantics}. Considering Figure~\ref{fig:pierre}, as $l$ is always
equal to $\delta$ when using Pierre or \friendloc, the total area of
the $9$ cells considered in proximity is $9 \delta^2$, while the area
of the circle is $\pi \delta^2$, which is almost $3$ times smaller.
This means that, assuming a uniform distribution of the users, using
Pierre or \friendloc the probability that a buddy reported as in
proximity is actually in proximity is about $1/3$.
On the contrary, in the protocols presented in this paper
the size of the granules is independent from the chosen
$\delta$. In our example, this means that when the value $l$ is
smaller than $\delta$, the region in which users are erroneously
reported in proximity becomes smaller (Figure~\ref{fig:hncapprox}).

\begin{figure*}[htb]
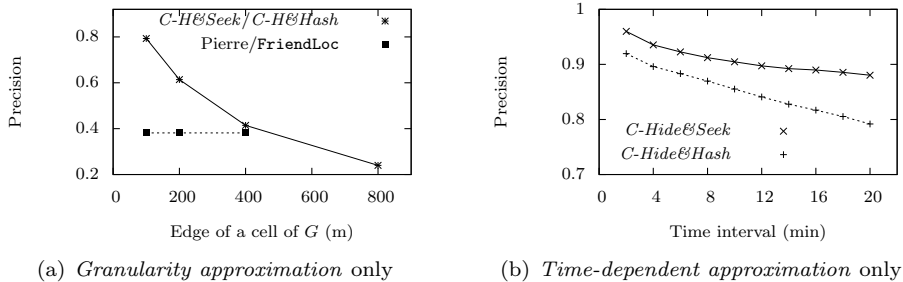

\centering
\subfigure[\emph{Granularity approximation} only]{
\label{fig:precision_edge}
\scalebox{0.8}{\input{fig/precision_edge_granappr}}
}
\subfigure[\emph{Time-dependent approximation} only]{
\label{fig:precision_time}
\scalebox{0.8}{\input{fig/precision_time}}
}
\caption{Evaluation of the impact of the approximations}
\label{fig:precision_approx}
\end{figure*}

Figure~\ref{fig:precision_edge} shows how the
\emph{granularity approximation} impacts on the service precision
for different values of the edge of granularity cells.
The metric we use for the measurement is the information retrieval
notion of \emph{precision}: the ratio between the number of correct
``in proximity'' answers over the total number of ``in proximity''
answers.
Intuitively, the precision measures the probability that
a buddy reported ``in proximity'' is actually in proximity.
Note that the analysis would be incomplete without considering the
notion of \emph{recall}: the ratio between the number of correct ``in
proximity'' answers over the sum of correct ``in proximity'' and
incorrect ``not in proximity'' answers.  Intuitively, the recall
measures the probability that a buddy actually in proximity is
reported ``in proximity''.  In this case, since we are considering the
\emph{minimum-distance} semantics (see Section~\ref{sub:qos}), the
\emph{granularity approximation} does not produce any incorrect ``not
in proximity'' answer, and hence the recall is equal to $1$.
When conducting this experiment, in order to exclude from the
evaluation the effects of the \emph{time-dependent approximation}, for
each buddy we used his current location as the last reported location.
Since Pierre and \friendloc do not consider $G$, their precision is
constant in the chart and, as expected, is below $0.4$.  On the
contrary, \chns and \chnh have a significantly better precision when
the edge of the cells of $G$ is small.  Intuitively, this is because
the area where a buddy is erroneously reported as in proximity is
smaller than $\delta$ (see Figure~\ref{fig:hncapprox}).
Figure~\ref{fig:precision_edge} also shows the precision when the edge
of a cell of $G$ is larger than $\delta$; The values are not reported
for Pierre and \friendloc since in this case they do not guarantee the
privacy requirements.

\begin{figure*}[htb]
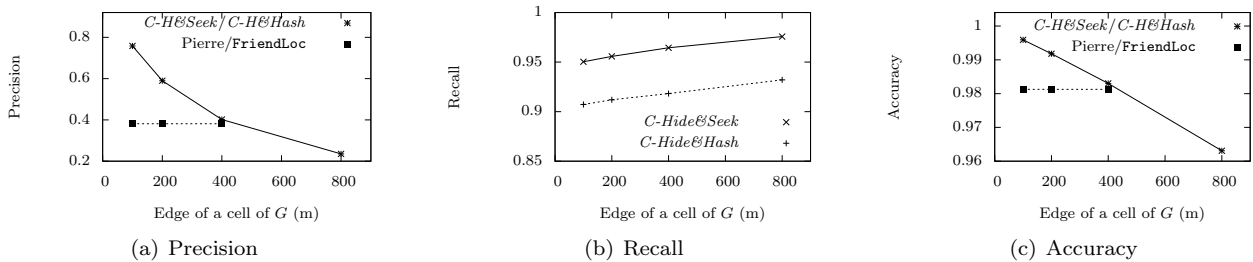

\centering
\subfigure[Precision]{
\label{fig:prec_delta}
\scalebox{0.72}{\input{fig/precision_edge}}
}
\subfigure[Recall]{
\label{fig:recall_edge}
\scalebox{0.72}{\input{fig/recall_edge}}
}
\subfigure[Accuracy]{
\label{fig:accuracy_edge}
\scalebox{0.72}{\input{fig/accuracy_edge}}
}
\caption{Evaluation of the quality of service (considering both approximations)}
\label{fig:servicePrecision}
\end{figure*}

Figure~\ref{fig:precision_time} shows the impact of the
\emph{time-dependent approximation}.  The chart shows the results for
our protocols only, as the other protocols proposed in the literature
are not exposed to this kind of approximation.  In order to exclude
from this evaluation the effects of the \emph{granularity
  approximation}, we performed these tests with the exact locations of
the users, instead of the generalized ones.  The chart shows, on the
$x$ axis, different lengths of the update interval and, on the $y$
axis, the precision of the \chns and \chnc protocols.  It can be
observed that \chns has better precision.  This is due to the fact
that \chnh always uses the location reported during the previous
update interval, while \hns uses the last location, that can be the
one reported during the current update interval or during the previous
one.
Since the \emph{time-dependent approximation} also introduces
incorrect ``not in proximity'' answers, we also measured the
recall. The corresponding chart is omitted as it is almost identical
to the one in Figure~\ref{fig:precision_time}.  For example, using
\chnh and an update interval of 4 minutes, the value of the precision
is $0.89$ and the recall is $0.88$.

The computation of the precision and recall under the
\emph{time-dependent approximation} confirms the intuition that using
long update intervals negatively impacts on the quality of service.
The choice of a value for the update interval should consider, in
addition to this approximation, the cost of performing a location
update.  In general, the optimal value can be identified based on
specific deployment scenarios.  Considering our movement data, we
chose $4$ minutes as a trade off value since it guarantees precision
higher than $0.9$ and sustainable system costs as detailed in
Section~\ref{sub:excosts}.  Our choice is consistent with similar
proximity services like, for example, Google Latitude that currently
requires location updates every $5$ minutes.
Figure~\ref{fig:servicePrecision} shows the analysis of the quality of
service considering both the \emph{granularity} and
\emph{time-dependent} approximations.  Figure~\ref{fig:prec_delta}
shows the precision of our two protocols compared with the precision
of Pierre and \friendloc.
We represent the precision of \chns and \chnh with a single curve
because the two protocols behave similarly.
For example, when the edge of a cell of $G$ is $200$m, the precision
of \chns and \chnh is $0.59$ and $0.57$, respectively, while it is
$0.61$ for both protocols when the \emph{time-dependent approximation}
is not considered.  This shows that this second type of approximation
does not have a significant impact.

Figure~\ref{fig:recall_edge} shows the recall of our protocols.  Note
that Pierre and \friendloc do not lead to incorrect ``not in
proximity'' answers, and hence their recall is equal to 1.  On the
contrary, our protocols can generate incorrect ``not in proximity''
answers due to the \emph{time-dependent approximation}.  This chart
shows that the recall of \chns and \chnh is always above $0.95$ and
$0.9$, respectively.
From Figure~\ref{fig:recall_edge} we can also observe that the recall increases for coarser granularities.
This is due to the fact that less incorrect ``not in proximity'' answers are returned
if a coarser granularity is used.
While this may appear unreasonable, the explanation is straightforward:
there is an incorrect ``not in proximity'' answer only when a buddy
is currently in proximity
(considering Figure~\ref{fig:hncapprox}, his location is in the dark gray area)
while the location used in the computation of the proximity is
outside the light gray area.
If a granularity is coarse, then the light gray area is large
and hence incorrect ``not in proximity'' are less frequent.

\begin{figure*}[ht!]
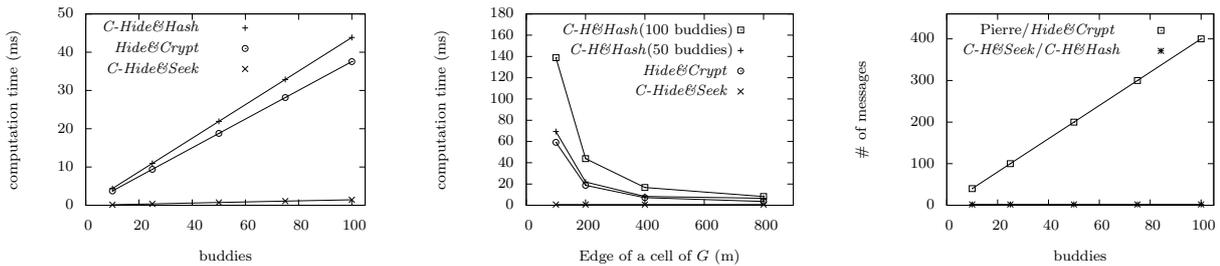
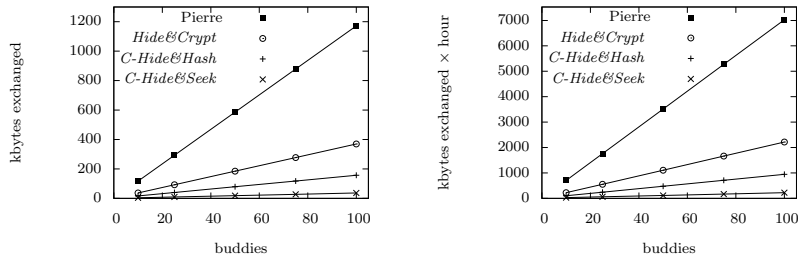

\centering
\subfigure[Computation time to issue a proximity request over number of buddies]{
\label{fig:pr_client_comp_buddies}
\scalebox{0.7}{\input{fig/pr_client_comp_buddies}}
}
\subfigure[Computation time to issue a proximity request over size of cells]{
\label{fig:pr_client_comp_edge}%
\scalebox{0.7}{\input{fig/pr_client_comp_edge}}%
}
\subfigure[Communication cost of a proximity request (\# of messages)]{
\label{fig:pr_client_commsg_buddies}
\scalebox{0.7}{\input{fig/pr_client_commsg_buddies}}
}
\subfigure[Communication cost of a proximity request (kbytes exchanged)]{
\label{fig:pr_client_comm_buddies}
\scalebox{0.7}{\input{fig/pr_client_comm_buddies}}
}
\subfigure[Hourly communication cost per user (location updates $+$ proximity requests)]{
\label{fig:service_cost}
\scalebox{0.7}{\input{fig/service_cost}}
}
\caption{Evaluation of the system costs}
\label{fig:qos}
\end{figure*}

Figure~\ref{fig:accuracy_edge} shows the \emph{accuracy} for each
considered protocol, i.e., the percentage of correct answers. Also in
this case, the accuracy of \chns and \chnh is represented with a
single curve, as the two protocols behave similarly.
Comparing this figure with Figure~\ref{fig:prec_delta}, we can observe
that the accuracy achieved by all the protocols is much higher than
the precision. This is due to the fact that this metric also considers
the correct ``not in proximity'' answers that are usually the most
frequent answers, since the proximity query area determined by the
distance threshold is usually much smaller than the entire space.
Figure~\ref{fig:accuracy_edge} shows that our protocols achieve better
accuracy than Pierre and \friendloc when the value of the edge of the
granularity cells is smaller than $\delta$.  In particular, for our
default values, the accuracy of both \chns and \chnh is higher than
$0.99$.

\subsection{Evaluation of the system costs}
\label{sub:excosts}

The second set of experiments evaluates the computation and
communication costs of the different protocols.
For the analysis of the Pierre protocol, we used the
NearbyFriend\footnote{\url{http://crysp.uwaterloo.ca/software/nearbyfriend/}}
application, developed by the same authors, which integrates the
Pierre protocol in a desktop IM application.

First, we consider the costs related to the location update
sub-protocol. This analysis does not apply to existing solutions as
location updates are only required by our centralized solutions.  As
analyzed in Section~\ref{sub:costs}, the temporal complexity of
computing a location update is constant in the number of buddies.  In
our implementation, the computation of each location update requires,
on the client side, about half of a millisecond for both the \chns and
the \chnc protocols.  Similarly, the communication cost is independent
from the number of buddies and the payload of each location update
message consists in few bytes.  Considering the overhead caused by the
XML encapsulation, the dimension of each location update is in the
order of a few hundred bytes.

The computation time needed to run a proximity request on the clients
is shown in Figure~\ref{fig:pr_client_comp_buddies} and
\ref{fig:pr_client_comp_edge}.  Note that the values reported in this
figure only consider the computation time required by the issuing
user.  Indeed, all the protocols require the SP (in case of
centralized services) or the other buddies (in case of distributed
services) to participate in the protocol, and hence to perform some
computation.  For example, in the case of \hnc and Pierre, the total
computation time of a user's buddies to answer a proximity request
issued by that user is about the same as the computation time required
to issue the request.  As observed in Section~\ref{sec:analysis}, the
computation time of a proximity request is linear in the number of
buddies.  Figure~\ref{fig:pr_client_comp_buddies} shows that \chnc
requires significantly more time with respect to \chns, especially
when the number of buddies is large.
However, the computation time of \chnh needed to issue a proximity
request for $100$ buddies is only about $40$ms on the mobile device we
used in our experiments.  The figure also shows that the computation
times of \chnh and \hnc are similar, with \hnc performing slightly
better.  This is due to the fact that in \hnc each of the $sMax$
indexes only needs to be encrypted, while in \chnh it also needs to be
hashed.

For what concerns other existing solutions, we did not implement the
Pierre protocol on our mobile device platform.  However, considering
the experimental results presented by the authors (see
\cite{ZhongPET07}), the computation time of a single proximity request
with a single buddy is more than $350$ms\footnote{It is unclear
  whether this result is obtained on a mobile device.}.  Since, for
\chnh, the computation time on a mobile device of a proximity request
with a single buddy is less than $0.4$ms, according to the data we
have, our solution is more than $800$ times more efficient than the
Pierre solution.

Figure~\ref{fig:pr_client_comp_edge} shows the impact of the
size of the cells of $G$ on the computation time of a proximity
request.  As expected, this parameter does not affect the computation
time of \chns that is actually negligible, while it clearly has an
impact on \chnc.  Intuitively, when the cells of $G$ are small, a
large number of indexes needs to be encrypted and hashed
and, in our experimental settings,
the computation cost may grow up to almost $140$ milliseconds
with cells having edge of $100$m, when considering $100$
buddies. Although 
the computation time grows
quadratically with the inverse of the edge of a cell, we believe that using
cells with edge smaller than $100$m would not justify the employment
of this privacy preserving technique. 
Indeed, a cell with an area smaller than $100\times100$m denotes a very low privacy requirement, while \chnc is preferable over \chns only when strong privacy is required.

Regarding the computation costs on the server side,
the complexity of a proximity request using \chnc
on the server side is similar to the one on the client side.
However, in our experiments we observed that our high-end desktop
machine is about 
$7$ times faster than the mobile client to execute these
operations. As a consequence, the computation for a single user having
$50$ buddies requires around $3$ms.  While we did not run scalability
tests on our server, this result suggests that, from the computational
point of view, even a single desktop machine can provide the service
to a large number of users.

Figures~\ref{fig:pr_client_commsg_buddies} and \ref{fig:pr_client_comm_buddies} show
the system communication cost of a proximity request issued by a user.
In Figure~\ref{fig:pr_client_commsg_buddies} we measure the number
of messages exchanged by the system for each proximity request.
It is easily seen that using a centralized protocol (i.e., \chns and \chnh),
only two messages need to be exchanged (one for the request and one for the response)
independently from the number of buddies the issuer has.
On the contrary, the decentralized protocols requires at least two messages for each buddy.
Moreover, in our implementation of the \hnc protocol,
each communication between two users needs to transit through the SP.
The same applies to the Pierre protocol, using the NearbyFriend implementation.
Consequently, at each location update, for each buddy, four messages transit in the system:
two between the issuer and the SP and two between the SP and the buddy.

Figure~\ref{fig:pr_client_comm_buddies} shows a comparison of the total amount of data
exchanged in the system for each proximity request.
Consistently with our analysis,
the communication cost grows linearly with the number of buddies
for both of our centralized protocols.
It is easily seen that this also applies to the other protocols.
The chart shows that NearbyFriend incurs in high communication costs.
The reason is that, each time a proximity request is issued,
a message of almost $3$KB is sent from the user to each of her buddies
and a message having a similar size is sent back in the reply.
We believe that this overhead is mostly given by the fact that NearbyFriend
needs all the communications between two users to be
encapsulated in a secure channel.
This is required because the Pierre protocol itself does not guarantee that any third party acquiring the messages
cannot derive location information about the users.
Since each message between two users
transits through the server,
the communication cost is almost $12$KB for each buddy.
The other decentralized solution we compare with, \hnc,
has better communication costs.
Indeed, each message is less than $1$KB, and hence
the cost is about $1/4$ if compared to Pierre.

Our centralized solutions are even more efficient.
This is due to the fact that only two messages need to be exchanged
between the user and the SP for each proximity request.
In case of \chnh, each message has the same dimension than in \hnc,
and hence, in this case, the communication cost is one half
with respect to \hnc, and about one order of magnitude less with respect to Pierre.
Finally, \chns, in addition to being a centralized solution,
also benefits from the fact that each message contains only a few hundred of bytes.
Consequently, this protocol is about $4$ times more efficient than \chnh.

In Figure~\ref{fig:service_cost} we evaluate the
communication cost of the continuous use of a proximity service with
our protocols.
As mentioned in Section~\ref{sub:exqos}, we consider that location updates
are issued every $4$ minutes.
Considering the results of our user survey,
we use $10$ minutes as the average frequency of proximity requests. 
The main difference of this figure with respect to
Figure~\ref{fig:pr_client_comm_buddies} is that it also considers the
communication costs derived by the location updates.  However, since
each location update costs less than $300$ bytes, and $15$ location
updates need to be issued in one hour, the total hourly cost for this
sub-protocol is about $4$KB, which is negligible with respect to the
communication cost of the proximity requests.  The figure also shows
that the centralized protocols require significantly less
communication than the decentralized ones.  In particular, \chns for
one hour requires around $100$KB when the user has $50$ online
buddies.  \chnh, on the other side, requires less than $500$KB per
hour for the same number of buddies.  We believe that this cost is
largely sustainable on a wireless broadband network (e.g., 3G), and
that, given the additional privacy with respect to curious buddies
achieved using \chnh, privacy concerned users may find this trade-off
attractive.

Our experimental evaluation also included the measurement of the cost
to distribute the private key (see Section~\ref{sec:implementation}).
Both the computation and communication costs are linear in the number
of buddies that need to receive the new key.  For a single buddy, the
computation time is about $7$ms, measured on the mobile device, while
the communication cost is less than $200$ bytes. An experiment of key
distribution to $50$ buddies, resulted in a computation time of $340$
ms, and a communication cost of less than $9$KB.

\subsection{Evaluation of the achieved privacy}
\label{sub:expriv}
In Section~\ref{sec:analysis} we proved that both of our protocols
guarantee the users' privacy requirements. 
We also observed that that \chnh provides more privacy than
what would be strictly necessary to
guarantee the requirements.
In this last set of experiments we evaluate how much additional privacy is provided by \chnh 
in terms of the size of the uncertainty region.
We recall that this is the area where a user $A$ is possibly located
as it can be computed by one of $A$'s buddies after
issuing a proximity request that returns $A$ as in proximity.

\begin{figure}[htb]%
\centering
\scalebox{1}{\input{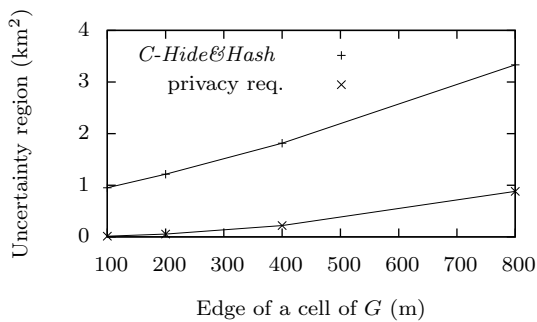}} %
\caption{Size of the uncertainty region.}
\label{fig:privacy_edge}
\end{figure}

Figure~\ref{fig:privacy_edge} shows that the privacy provided by \chnh is always
significantly larger than the privacy requirement, and it grows for coarser granularities $G$.
Intuitively, with \chnh, the uncertainty region corresponds to the union of the light and dark gray areas
represented in Figure~\ref{fig:hncapprox}.
Consequently, as the size of the cells of $G$ decreases,
the size of the light gray area tends to zero,
and the uncertainty region becomes closer and closer to the
dark gray area only.
This means that the privacy provided by \chnh is at least
$\pi \delta^2$ even when the user requires her location to be
obfuscated in a smaller area.

\section{Discussion and conclusions}
\label{sec:conc}
We presented a comprehensive study of the location privacy problem associated with the use of a proximity service  
as an important component of any geo-social network.
We illustrated two new protocols to compute users' proximity that take
advantage of the presence of a third party to reduce the computation
and communication costs with respect to decentralized solutions.
We formally proved that the service provider acting as the third party, by
running the protocol, cannot acquire any new location information about
the users, not even in presence of a-priori knowledge of users' locations. 
We also showed that each user can have full control of the location
information acquired by her buddies. Extensive experimental work and a complete implementation illustrate the benefits of the proposed solutions with respect to existing ones, as well as their actual applicability.  

The two centralized solutions we propose require each user to share
keys with her buddies, and hence are not the most appropriate to be used in
a ``query driven'' service
(e.g., finding people meeting certain criteria).
The decentralized versions of the two presented protocols are
more suitable in this case \cite{MascettiMDM09}.

An interesting direction we plan to investigate is to extend the
adversary models we considered in this paper to include 
not only (atemporal) a-priori location knowledge, but also \emph{time-dependent} location knowledge. 
This would model not only a-priori knowledge about velocity, that our solutions can already deal with, but also a-priori probabilistic \emph{proximity} information.  It is still unclear if the proposed protocols, with
appropriate location update strategies, similar to those discussed in Section~\ref{sub:velocity}, need to be modified in order to be proven
privacy-preserving according to our definitions.

An interesting extension of our protocols is to allow users to specify
different privacy preferences with respect to different groups of buddies.  This
is not difficult, but it exposes the users to dangerous collusion attacks if
further constraints are not imposed.  The presented protocols are not
subject to buddies' collusion attacks since each user defines the same
granularity as privacy preference with respect to all of her buddies.
If this is not the case, a user $A$, by assigning two different
granularities with respect to buddies $B$ and $C$ to reflect her
different level of trust, would expect that if $B$ and $C$ collude the
lowest privacy requirement among the two is preserved. However, an adversary could actually intersect the uncertainty regions and potentially violate both privacy requirements. In order for
our protocols to defend against such a collusion, some relationships need to be
imposed on the granularities used in the system. While details are out
of the scope of this paper, intuitively, granules from different
granularities should never partially overlap.  For example, using
hierarchical grids as granularities would be a sufficient condition.

Finally, our solution is limited to location privacy and it does not
enforce anonymity since the participating buddies often know each
other. We do not exclude that for some proximity services anonymity
would be desirable in order, for example, to hide to the SP the real identity of buddies or each user's list of buddies. However, since the
techniques proposed in this paper guarantee that no location
information is disclosed to the SP, it would not be difficult to adapt
them to provide anonymity with respect to the SP by applying, for
example, existing anonymization techniques for standard (i.e., non
spatio-temporal) datasets.

\section*{Acknowledgments}
The authors would like to thank the anonymous reviewers for their insightful comments and useful suggestions. 
This work was partially supported by Italian MIUR under grants PRIN-2007F9437X and InterLink II04C0EC1D, and by the National Science Foundation under grant CNS-0716567.

\bibliographystyle{plain}

\newpage
\appendix

\section{Proofs of formal results}
\label{sec:proofs}
\subsection{Proof of Lemma \ref{le:location_message_independence}}
\begin{proof}
The sought after independence intuitively means that whether an update/request is sent to SP by a user $A$ is not related to where the user is located. Formally, by the definition of conditional probability, we have 
\begin{eqnarray*}
P(ur(A) | loc_{A}, pri_A)
\end{eqnarray*}
\vspace*{-0.3in}
\begin{eqnarray*}
 & = & P(ur(A), loc_A| pri_A)/P(loc_A|pri_A)\\
 & = & (P(ur(A)|pri_A)*P(loc_A|pri_A))/P(loc_A|pri_A)\\
 & = & P(ur(A)|pri_A). 
\end{eqnarray*}
The second equality is due to the protocol, in which an update/request is sent at fixed time intervals for each user {\em independent} of the user's location. Hence, the lemma follows.
\end{proof}

\subsection{Proof of Lemma \ref{le:message_independence}}
\begin{proof}
All we need is
$$P(M_1|M_2,loc_A, pri_{A})=P(M_1|loc_A, pri_{A}),$$
i.e., the knowledge of the messages in $M_2$ does not have any impact on the probability of messages in $M_1$. But this follows the perfect secrecy assumption and the use of keystreams in our protocol.
\end{proof}

\subsection{Proof of Theorem \ref{th:hnsSP}}
\begin{proof}
We prove the theorem by showing that
for each set $M$ of messages exchanged
during the protocol, we have
$P(post_{A})=P(pri_{A}).$
That is, the messages $M$ do not change the $SP$'s knowledge of $A$'s location.
By assumption of the theorem, $P(post_{A})=P(loc_{A} | M, pri_{A})$ as the only knowledge is $M$ and $pri_{A}$. The knowledge that $loc_A\in g_A$ is useless as we assume in this case that $g_A$ is the whole spatial domain.
By the definition of conditional probability, we have\\
\hspace*{0.1in}$P(loc_{A} | M, pri_{A})=$\\
\hspace*{0.1in}$
P(M | loc_{A}, pri_{A})*P(loc_{A}| pri_{A})/P(M|pri_{A}).
$\\
It now suffices to show 
\begin{eqnarray}
P(M|loc_{A}, pri_{A})=P(M|pri_{A}).
\label{eq:hnsproof}
\end{eqnarray}
Intuitively, Equation~\ref{eq:hnsproof} says that the messages $M$ are independent of the location of $A$. This follows from two observations: the first is that the issuance of messages does not depend on the location of $A$ by Lemma~\ref{le:location_message_independence} and the second is that the (encrypted) messages are independent of the content of the messages by Lemma~\ref{le:message_independence}.
More formally, assume
$$M  =   m_{1}, \ldots, m_{n}.$$
Let $ur(M)$ be the messages of the form 
$$ur(m_{1}),\ldots, ur(m_{n}),$$
where $ur(m_{i})$ is ``an update/request is sent by user $B_{i}$''. That is, $ur(m_{i})$ disregards the encrypted part of the message but only says that a message is sent and by whom.
By perfect secrecy assumption, the probability of a particular (single) message is the same as any other (single) message that differs only in the encrypted part, and hence the same as the probability of $ur(m_i)$. Consider the case of two messages in $M$, i.e., $n=2$.
Now we have:
\begin{eqnarray*}
&P&(M|loc_A, pri_{A})\\
& = & P(m_1, m_2|loc_A, pri_{A})  \mbox{}\\
	& = & P(m_1|m_2, loc_A, pri_{A})*P(m_2|loc_A, pri_{A})  \mbox{}\\
	& = & P(m_1|loc_A, pri_{A})*P(m_2|loc_A, pri_{A}) \mbox{ {\rm by Lemma~\ref{le:message_independence}} }\\
	& = & P(ur(m_1)|loc_A, pri_{A})*P(ur(m_2)|loc_A, pri_{A})  \\
	&   & \mbox{\hspace*{0.2in} {\rm by the above discussion} }\\
	& = & P(ur(m_1),ur(m_2)|loc_A, pri_{A}) \mbox{ {\rm by Lemma~\ref{le:message_independence}} }\\
	& = & P(ur(M)|loc_A, pri_{A})
\end{eqnarray*}
The above can be extended to $n$ messages in $M$ and also to show the equation $P(M|pri_{A})=P(ur(M)|pri_{A})$. Hence,
\begin{eqnarray*}
&P&(M|loc_{A}, pri_{A})\\
& = & P(ur(M)|loc_{A}, pri_{A})\\
& =&  P(ur(M)|pri_{A}) \mbox{ by Lemma~\ref{le:location_message_independence}}\\
& = & P(M|pri_{A})
\end{eqnarray*}
and the thesis is established.
\end{proof}

\subsection{Proof of Theorem \ref{th:hnsbuddies}}
\begin{proof}
Given a buddy $B$, we prove the theorem by showing that
for each set $M$ of messages exchanged
during the protocol, we have
\[P(loc_{A} | M, pri_{A}, loc_{A}\in g_A) = P(loc_{A} | pri_{A}, loc_{A}\in g_A), 
\]
where $A$ is another user, and $g_A$ is the location information that is encrypted in the messages of $A$ with the key shared between $A$ and $B$. In other words, we want to show that $B$ will not acquire more location information about $A$ through the messages other than what $B$ already knows. Intuitively, this is true since the location information revealed by $A$ is only at the granule level, but not {\em where} within the granule.

The formal proof is the same as for Theorem~\ref{th:hnsSP} but with the following two changes: (1) $ur(m)$ represents that request was sent from the granule included in the message if the message is intended to $B$; otherwise, it is the same as before. (2) $loc_A\in g_A$ is included in $pri_A$, or equivalently we replace each occurrence of $pri_A$ with ``$loc_A\in g_A, pri_A$''. Let us now examine the steps in the proof of Theorem~\ref{th:hnsSP}.

Lemma~\ref{le:location_message_independence} still holds since updates/requests are sent regardless of locations if the user who sent the message is $C\ne A$. If $C=A$, then the $ur(A)$ gives the location (the granule) where the message is sent. In this case, the location is totally dependent on the given information of $loc_A$, $loc_A\in g_A$ and $pri_A$. Note that $l$ is an index of a granule, any information contained in $loc_A$ and $pri_A$ below the granule level is not relevant to the probability of a message.

For Lemma~\ref{le:message_independence}, the content in $M_2$ still does not have any impact on the content in $M_1$ even when $B$ can decrypt the messages intended to him as there is no information (from $pri_A$, $loc_A$, and $loc_A\in g_A$) that restricts any possible content in $M_1$, so the conditional probability of $M_1$ does not change regardless the existence of $M_2$. 

For the discussion regarding the probability of $m_i$ and $ur(m_i)$, 
with the addition of $loc_A\in g_A$, we still have that the conditional probability of $m_i$ being the same as that of $ur(m_i)$. Indeed, assume
\[m_i=\langle C, ui, E_{K^{ui}}(l)\rangle.\]
If $C\ne A$, then all messages of the type have the same probability with or without knowing $A$'s location since $C$'s location information is not assumed in the conditional probability. This case is exactly the same as for the SP and the conditional probability of $m_i$ is the same as that of $ur(m_i)$. If $C=A$, since $B$ can decrypt the message, hence knowing the location $l$ in the message, this location $l$ (an index value of a granule in $G_A$) needs to be consistent with the location knowledge in $loc_A$ and $pri_A$: if it is not consistent, then the probability of the message is zero; otherwise, the probability is totally dependent on the probability of $A$ being in $G_A(l)$ given $loc_A$, $loc_A\in g_A$, and $pri_A$. But the same can be said about $ur(m_i)$ (which says that a message was sent at the given location), i.e., the probability of $ur(m_i)$ depends totally on $loc_A$, $loc_A\in g_A$, and $pri_A$. Therefore, $m_i$ and $ur(m_i)$ have the same conditional probability. By the same reasoning as in the proof of Theorem~\ref{th:hnsSP}, $ur(M)$ has the same conditional probability as $M$.

With all the above discussions, the theorem is established.
\end{proof}

\subsection{Proof of Theorem \ref{th:hncSP}}
\begin{proof}
The proof follows the same style of that for Theorem~\ref{th:hnsSP}. That is, we show $P(M|loc_A, pri_A)=P(M|pri_A)$, i.e., the location of $A$ does not change the probability of messages $M$ conditioned on $pri_A$. Like for Theorem~\ref{th:hnsbuddies}, we examine the proof steps of Theorem~\ref{th:hnsSP} for the purpose of the current thesis. Lemmas \ref{le:location_message_independence} and \ref{le:message_independence} both hold due to the use of hashing function that displays stronger secrecy than encryption. The important difference is the discussion of the conditional probabilities of $m$ and $ur(m)$. If $m$ is an update, then the same applies as in the proof of Theorem~\ref{th:hnsSP}. The difference is when $m$ is a proximity request. In this case, the message contains multiple components. The critical step is to show that all such messages have the same conditional probability (to the SP) and hence the same as the conditional probability of $ur(m)$. This is not difficult since the location information in the condition is opaque to the SP. This opaqueness is given by two facts. The first is that the number of components in the message is the same regardless of the location information. The second is that the indexes of the granules and the ``padding'' ($S''$ in the protocol) in the message components are hashed and hence to the SP all possible granule indexes are equally possible in the encrypted (by $K_1$ in the protocol) message. (Here, hashing before encryption with $K_1$ is important as the adversary cannot attack using known pattern of the plaintext.) The above observations lead to the thesis of this theorem.
\end{proof}

\subsection{Proof of Theorem \ref{th:hncbuddies}}
\begin{proof}
Intuitively, to the buddies, the \chnc is much stronger than \chns since buddies only share a hashing function and the buddies location information is encrypted by a random key (generated by the SP) before sending to the requesting user $B$. Formally, the proof follows the same style as that for Theorem~\ref{th:hnsbuddies}. The only difference is what it means when a message is ``consistent'' with the location knowledge. In this case, from $B$'s perspective, we need to define $ur(m)$ to be the binary random variable that ``the user is in one of the requesting granules or not'' for the message sent back from the SP (as the reply to a proximity request from $B$).
After $B$ requesting proximity, $B$ will receive a message from the SP with the encrypted hash value of $A$'s location (in addition to the ``kick back'' from the SP in the form of encrypted values that $B$ sent to the SP). Even though $B$ and $A$ shares the hash function, $B$ does not know the encryption key which is randomly generated by the SP ($K_2$ in the protocol). Therefore, this value is probabilistically independent of the location of $A$. In this case, based on the protocol, the only information $B$ obtains is whether $A$ is in a granule among the ones given by $B$. This needs to be consistent with the location information contained in $loc_A$ and $pri_A$. If not, then the probability of this message is zero, and otherwise the probability is totally dependent on $loc_A$ and $pri_A$ as no other information is available. The thesis follows the above discussions in the same style as the proof of Theorem~\ref{th:hnsbuddies}.
\end{proof}

\end{document}